\documentclass[amsmath,superscriptaddress,showpacs,aps,prb,twocolumn]{revtex4-1}
\usepackage[colorlinks,linkcolor=blue,anchorcolor=blue,citecolor=blue,urlcolor=blue]{hyperref}
\usepackage{amsmath}
\usepackage{amssymb}
\usepackage{graphicx}
\usepackage{color}
\usepackage{bm}

\begin{document}
\title{Reentrant Localized Bulk and Localized-Extended Edge in Quasiperiodic Non-Hermitian Systems}

\author{Gang-Feng Guo}
\affiliation{Lanzhou Center for Theoretical Physics, Key Laboratory of Theoretical Physics of Gansu Province, Lanzhou University, Lanzhou $730000$, China}

\author{Xi-Xi Bao}
\affiliation{Lanzhou Center for Theoretical Physics, Key Laboratory of Theoretical Physics of Gansu Province, Lanzhou University, Lanzhou $730000$, China}

\author{Lei Tan}
\email{tanlei@lzu.edu.cn}
\affiliation{Lanzhou Center for Theoretical Physics, Key Laboratory of Theoretical Physics of Gansu Province, Lanzhou University, Lanzhou $730000$, China}
\affiliation{Key Laboratory for Magnetism and Magnetic Materials of the Ministry of Education, Lanzhou University, Lanzhou $730000$, China}

\begin{abstract}

The localization is one of the active and fundamental research in topology physics. Based on a generalized Su-Schrieffer-Heeger model with the quasiperiodic non-Hermitian emerging at the off-diagonal location, we propose a novel systematic method to analyze the localization behaviors for the bulk and the edge, respectively. For the bulk, it can be found that it undergoes an extended-coexisting-localized-coexisting-localized transition induced by the quasidisorder and non-Hermiticity. While for the edge state, it can be broken and recovered with the increase of the quasidisorder strength, and its localized transition is synchronous exactly with the topological phase transition. In addition, the inverse participation ratio of the edge state oscillates with an increase of the disorder strength. Finally, numerical results elucidate that the derivative of the normalized participation ratio exhibits an enormous discontinuity at the localized transition point. Here, our results not only demonstrate the diversity of localization properties of bulk and edge state, but also may provide an extension of the ordinary method for investigating the localization.
\end{abstract}

\maketitle
\section{INTRODUCTION}

Topological insulators have become a burgeoning research arena in recent years, in which the central concept is the topological state predicted by the topological invariant [\onlinecite{Bansil1, Chiu2, As3, AND4, Shen5}]. Within the conventional framework, the topological state is protected at the boundaries, and correspondingly, the bulk eigenstates suffuse over in the system simply [\onlinecite{As3}]. However, the localization behaviors mentioned above will become different as the non-Hermiticity is presented in the system [\onlinecite{16YK, 17MA, Wu171, He172, 18FT, 17, 18, 19, 20, 22, 23, 24, 27, 28, 29, 33, 34, 35, 37, 38, 39, 40, 41, 42, 44, 45, 47, 48, 49, 50}]. All bulk eigenstates are localized at the boundaries, which is dubbed as the non-Hermitian skin effect [\onlinecite{Le6, Kun7, Ye8, Yao9, Yok10, Zha11, Yan12, Li13, Gan14, Bao15}] and characterized by the spectral windings [\onlinecite{Gon16, Ana17, She18, Ley20, Denn201, Yin21, Kaw23, Oku24}]. Inversely, the topological edge states can be delocalized at the boundaries [\onlinecite{Zha25, Lon26, Zhu27}].

Quasidisorder is another basic quantity in condensed matter physics. It is known that the Anderson localization is induced by the quasidisoder, i.e., wave-functions are accommodated at any location of the system, not just at the boundaries, which distinguishes with the effect of the non-Hermiticity essentially. Refs. [\onlinecite{Che28, Jia29, Wan30, Tan31, Aga32, Zen33, Liu34, Ton35, Zhai36, Cai37, Long38, Zha40, Zhai41, Wang42, Zhao43, Long44, Zhou45, Jian46, Hou47, Zhang49}] show that the combination of the Non-Hermiticity and the quasidisorder can endow the system with more peculiar and unique phenomena. Such as, Ref. [\onlinecite{Zen33}] showed that the zero-energy mode locates only at one edge. Ref. [\onlinecite{Liu34}] exhibited that the localized transition can be connected with the $P$$T$-symmetry. In order to describe the localization phenomenon of the system accurately, the concepts of the inverse participation ratio and normalized participation ratio have been introduced [\onlinecite{Li50}, \onlinecite{Xiao51}]. However, systems combining the non-Hermitian and quasiperiodic have been generally analyzed as a whole rather than individually, and more nuanced analyses have not been carried out unfortunately. Here, more interesting hidden localization phenomena can be uncovered by strictly dividing the system into different segments (bulk vs. edge) within the framework of the sublattice symmetry.

Considering a generalized Su-Schrieffer-Heeger (SSH) system in which the quasiperiodic non-Hermitian locates at the off-diagonal enabling the existence of the winding number and the topological edge state exactly. Numerical calculations show that the bulk can undergo an extended-coexisting-localized-coexisting-localized transition, while the edge state experiences an appearance-disappearance-recurrence localized transformation accompanied by the topological phase transition. Additionally, a steep discontinuity for the derivative of the normalized participation ratio appears at the localized transformation point. Our work provides a comprehensive and finicky perspective to understand the phenomenon of the localization of the system.

The paper is organized as follows: A paradigm and the theoretical framework are constructed in Sec. \ref{II}, where the topological invariant in real space, the inverse participation ratio and normalized participation ratio are introduced. Sec. \ref{III} focuses on the localized transition and the topological phase transition of the edge state simultaneously. The derivative of the normalized participation ratio at the localized transition point is also discussed. In Sec. \ref{IV}, the bulk reentrant localization phenomena are investigated. Finally, the conclusions are found in Sec. \ref{V}.

\section{MODEL AND THEORY}\label{II}

We consider a one-dimensional non-Hermitian quasidisorder system. The Hamiltonian reads

\begin{flalign}
H=&t_{1}\sum_{n=1}^N({C^\dag}_{A,n}{C}_{B,n}+H.c.)+t_{2}\sum\limits_{n=1}^{N-1}({C^\dag}_{A,n+1}{C}_{B,n}+H.c.)\nonumber\\
&+W_{1}\sum\limits_{n=1}^{[\frac{N}{2}]}({C^\dag}_{A,2n}{C}_{B,2n-1}+H.c.)\cos(2\pi\beta n+i\gamma)\nonumber\\
&+W_{2}\sum\limits_{n=1}^{[\frac{N+1}{2}]-1}({C^\dag}_{A,2n+1}{C}_{B,2n}+H.c.)\cos(2\pi\beta n+i\gamma),\label{2.1}
\end{flalign}
where the first two terms represent the SSH model. $t_{1}$ and $t_{2}$ characterize the intracell and intercell hopping energies. [ ] is the integer ceiling function. $\gamma$ refers to the Non-Hermitian parameter. $\beta=\frac{\sqrt5 - 1}{2}$ shows the quasiperiodicity. $W_{1}$ and $W_{2}$ represent the quasidisorder strengths. The quasidisorders arise at the off-diagonal location, which endows the Hamiltonian the sublattice symmetry: $CHC^{-1}=-H$, with the chiral operator $C=diag(1,-1,1,-1,...,1,-1)$. Hence, the topological properties of our system undoubtedly can be revealed by the winding number and the corresponding topological zero-energy edge modes. In the real space, the winding number is defined as [\onlinecite{Song52}]

\begin{flalign}
\mu=\frac{1}{2 L^{'}}Tr{'}(CQ[Q,X]),\label{2.2}
\end{flalign}
where $X$ is the coordinate operator and $Q=\sum\limits_{n}(|\Psi^{(n)R}\rangle \langle\Psi^{(n)L}|-C|\Psi^{(n)R}\rangle \langle\Psi^{(n)L}|C^{-1})$. $|\Psi^{(n)R}\rangle$ and $|\Psi^{(n)L}\rangle$ satisfy $H|\Psi^{(n)R}\rangle=E_{n}|\Psi^{(n)R}\rangle$ and $H^{\dag}|\Psi^{(n)L}\rangle=E_{n}^{*}|\Psi^{(n)L}\rangle$, respectively. The length of the chain $L=2N$ will be divided into three intervals with length $l$, $L^{'}$, $l$ [i.e., $2l+L^{'}=L$], and $Tr^{'}$ only means the trace over the middle interval with length $L^{'}$. To quantize $\mu$ to an integer, both $L$ and $l$ should be sufficiently large.

In general, the localization properties of the wave functions can be captured fully using the inverse participation ratio (IPR) and the normalized participation ratio (NPR) in the large $L$ limit [\onlinecite{Li50}, \onlinecite{Xiao51}]. Notice that $\big||\Psi^{(n)R}\rangle \big| \neq \big||\Psi^{(n)L}\rangle\big|$ ($H\neq H^{+}$) as the system contains the non-Hermiticity. The localized properties of the wave function $|\Psi^{(n)R}\rangle$ should be organized by

\begin{flalign}
IPR^{|\Psi^{n}\rangle}=\frac{\sum\limits_{i=1}^L \left| \langle \Psi_{i}^{(n)R} |\Psi_{i}^{(n)R}\rangle\right|^{2}}{\Big[\sum\limits_{i=1}^L \left| \langle \Psi_{i}^{(n)R} |\Psi_{i}^{(n)R}\rangle\right|\Big]^{2}},\label{2.3}
\end{flalign}

and

\begin{equation}
NPR^{|\Psi^{n}\rangle}=(L\sum\limits_{i=1}^L \left| \langle \Psi_{i}^{(n)R} |\Psi_{i}^{(n)R}\rangle\right|^{2})^{-1}.\label{2.4}
\end{equation}

In our work, the localization behaviors of the bulk and edge are needed to be discussed separately. Then, the quantities characterized the bulk should be defined as
\begin{equation}
IPR^{Bulk}=\frac{1}{L-2}\sum\limits_{n=1}^{L-2} IPR^{|\Psi^{n}\rangle}, \label{2.5}
\end{equation}

\begin{equation}
NPR^{Bulk}=\frac{1}{L-2}\sum\limits_{n=1}^{L-2} NPR^{|\Psi^{n}\rangle}, \label{2.6}
\end{equation}
where $L$ stands for the length of the chain and $2$ is the number of lowest absolute energies values which equals the number of the topological zero-energy edge states exactly for the nontrivial case. Similar to $IPR$ and $NPR$ [\onlinecite{Li50}, \onlinecite{Xiao51}], $IPR^{Bulk}=0$ $(\neq0)$ and $NPR^{Bulk}\neq0$ $(=0)$ indicate the extended (localized) behaviors of the bulk eigenstates. In addition, $IPR^{Bulk}$ and $NPR^{Bulk}$ both are nonzero corresponding to the case that some eigenstates are extended while some others are localized. For simplicity, $IPR^{Bulk}$ and $NPR^{Bulk}$ are abbreviated as $IPR^{B}$ and $NPR^{B}$.

Similarly, the localization phenomena of the state with the lowest absolute energy can be described by $IPR^{Edge}$ and $NPR^{Edge}$ (shorthand by $IPR^{0}$ and $NPR^{0}$), i.e.,

\begin{equation}
IPR^{0}=\frac{1}{2}\sum\limits_{n=1}^{2} IPR^{|\Psi^{n}\rangle}, \label{2.7}
\end{equation}

\begin{equation}
NPR^{0}=\frac{1}{2}\sum\limits_{n=1}^{2} NPR^{|\Psi^{n}\rangle}. \label{2.8}
\end{equation}

On the novel concepts mentioned above, we will explore the topology and localization properties of the bulk and edge in the following sections.

\section{The Localized-Extended Edge}\label{III}

\begin{figure}
\includegraphics[width=4.2cm,height=3.8cm]{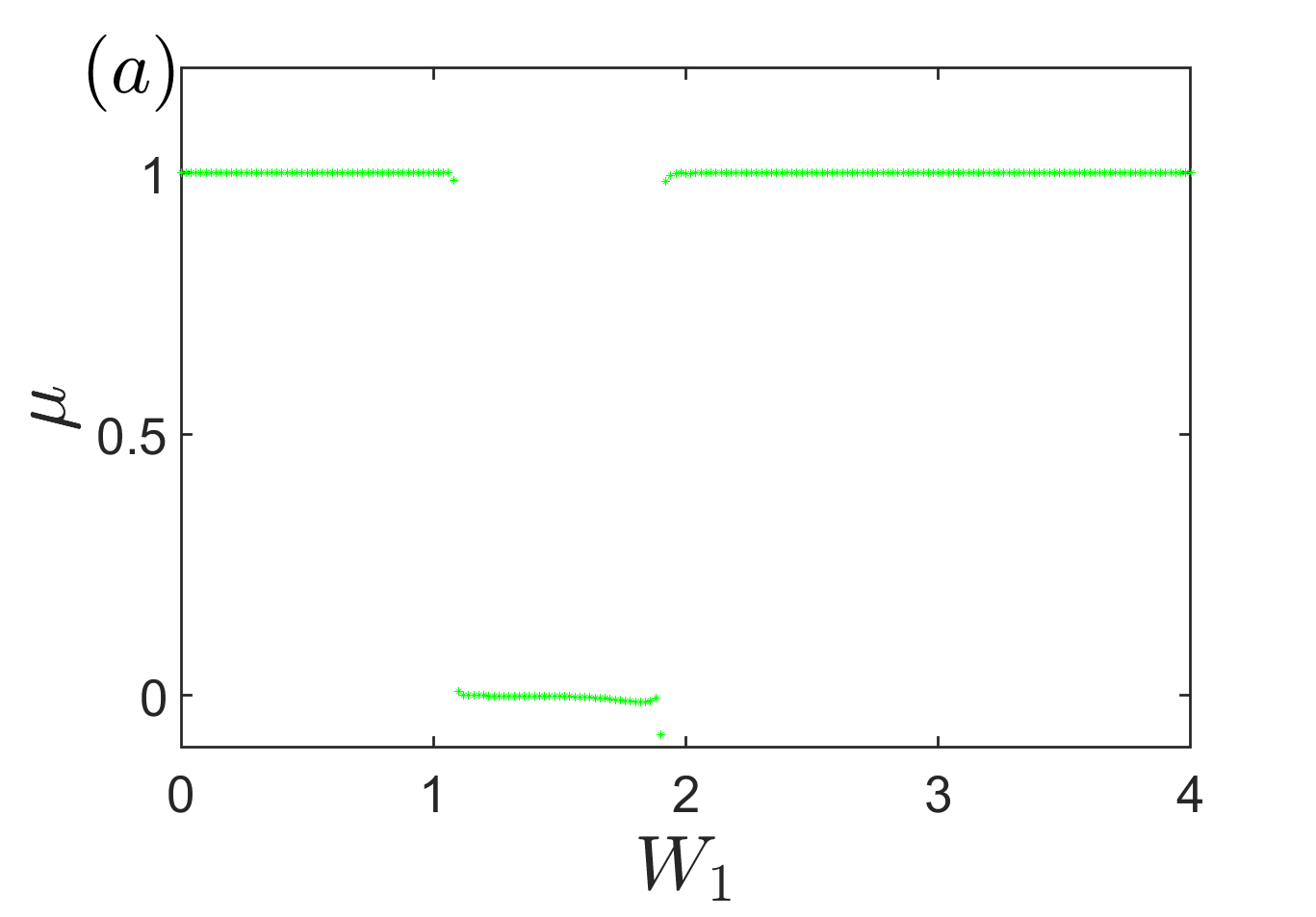}
\includegraphics[width=4.2cm,height=3.8cm]{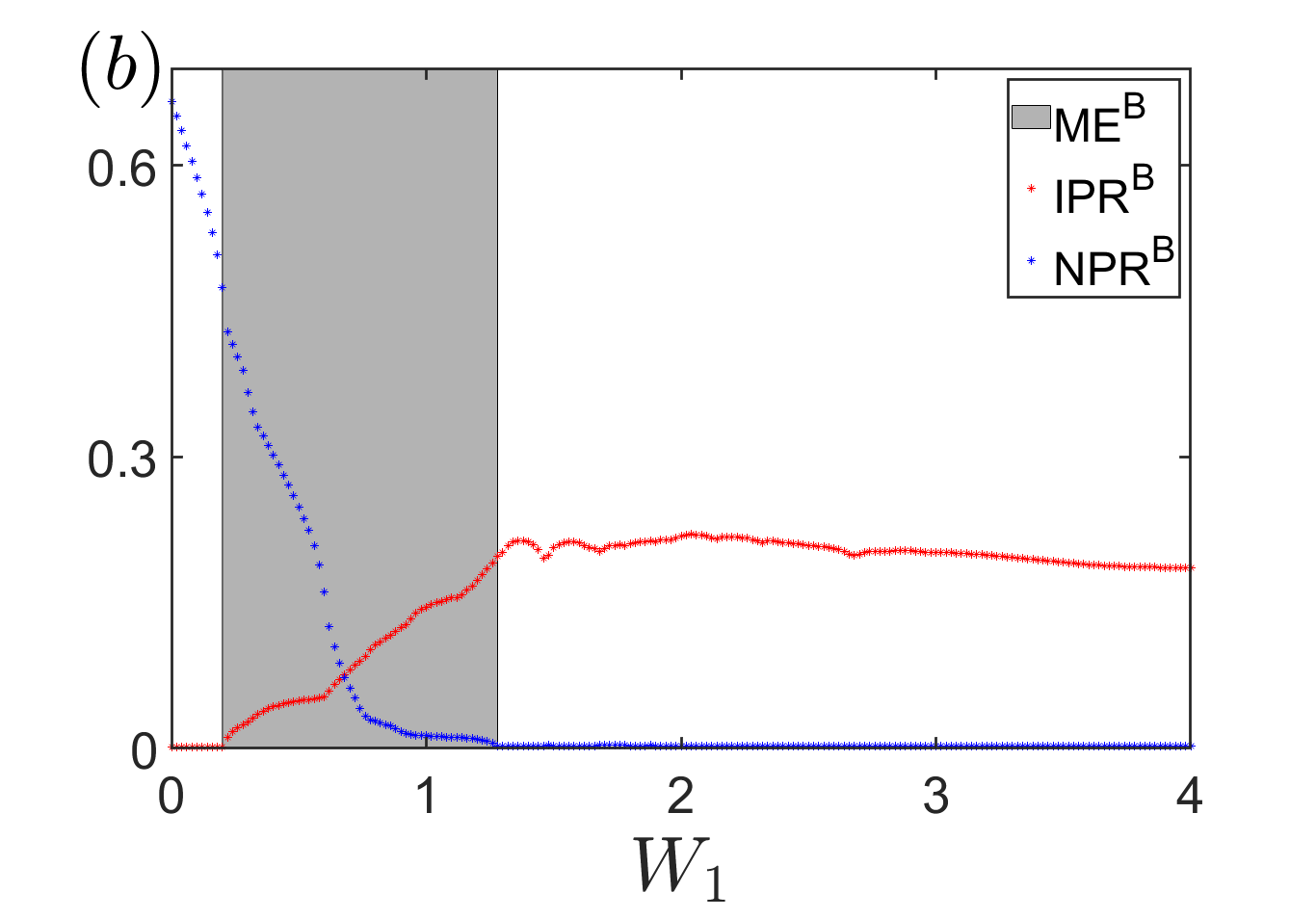}

\includegraphics[width=4.2cm,height=3.8cm]{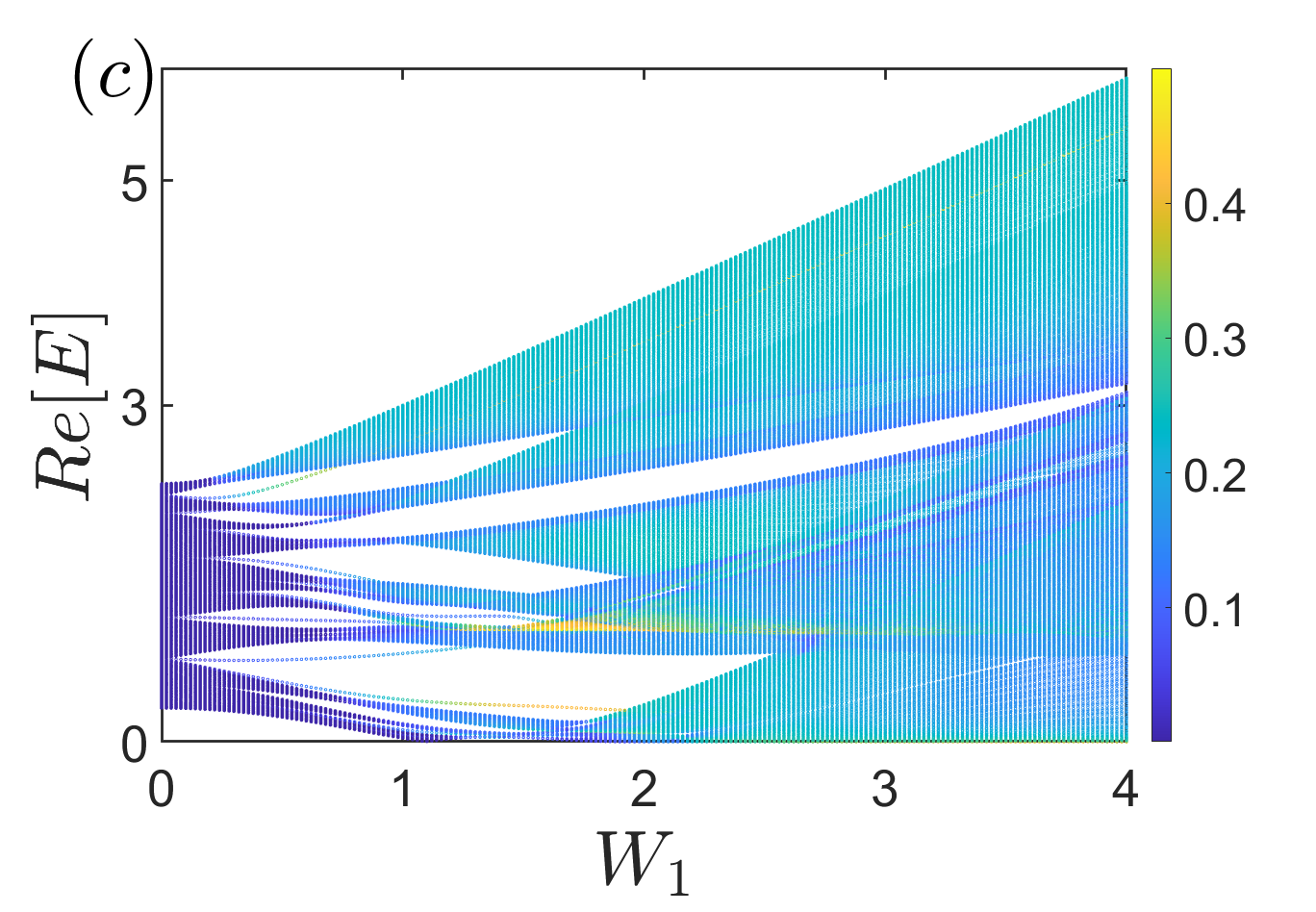}
\includegraphics[width=4.2cm,height=3.8cm]{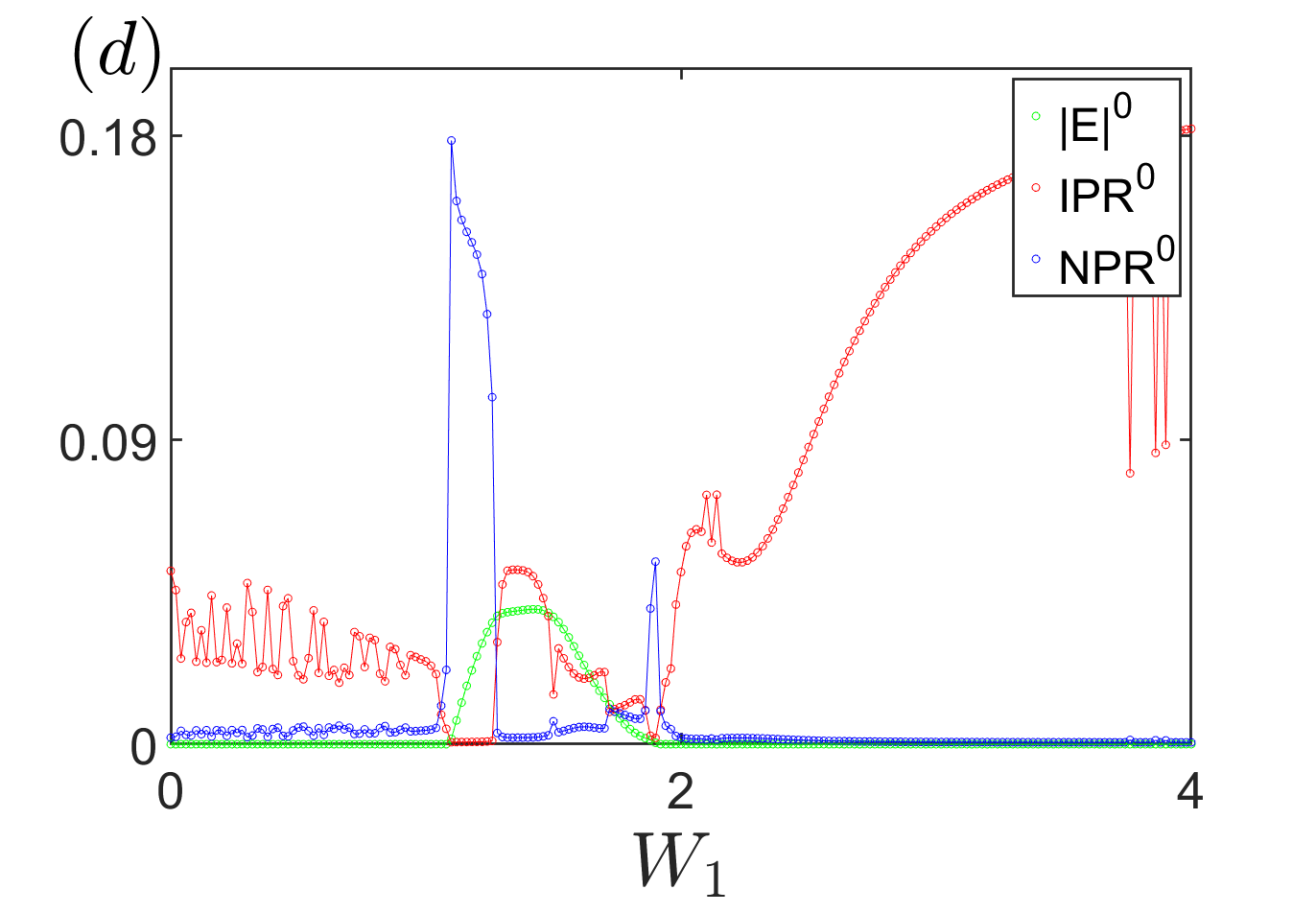}

\includegraphics[width=4.2cm,height=3.8cm]{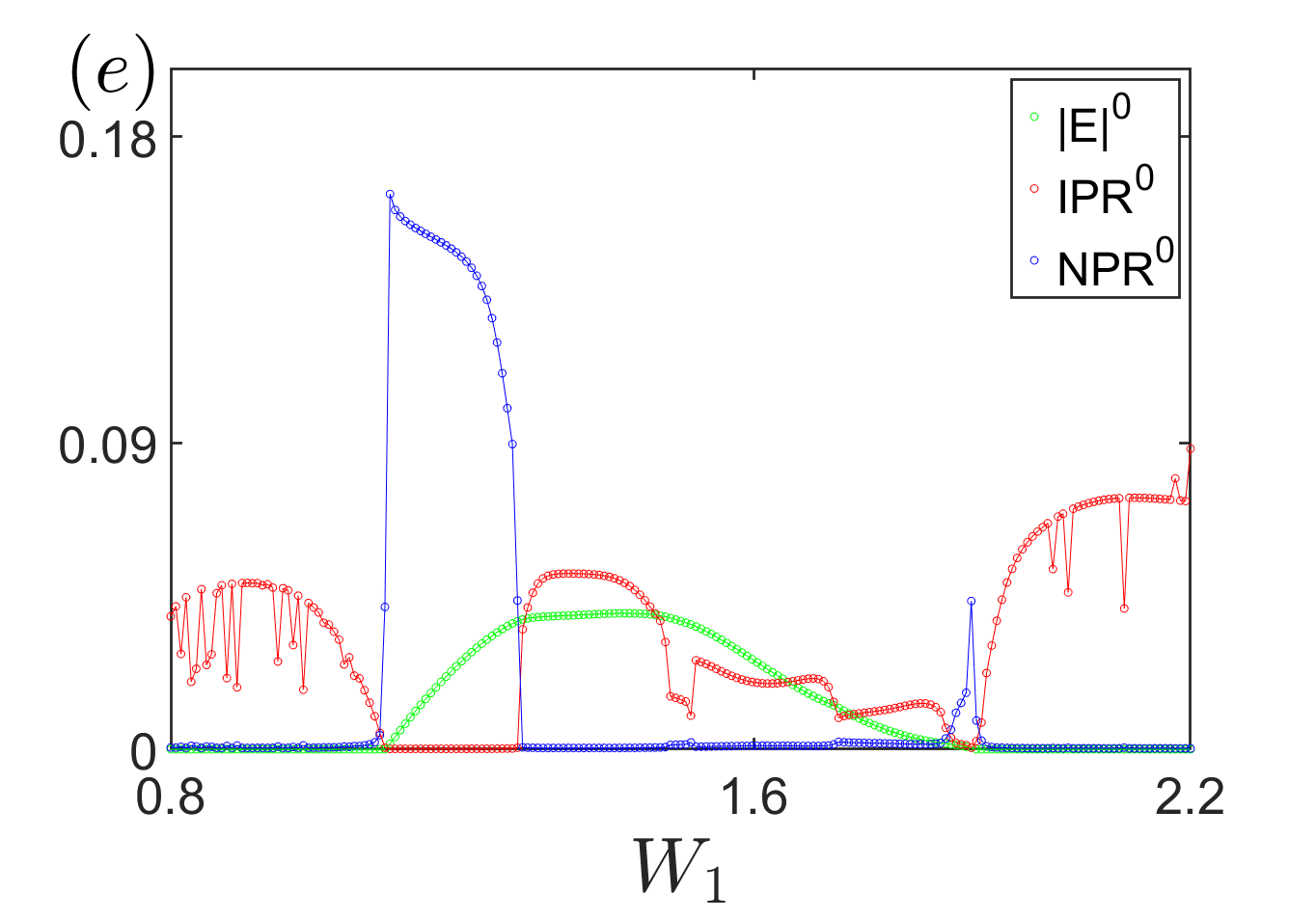}
\includegraphics[width=4.2cm,height=3.8cm]{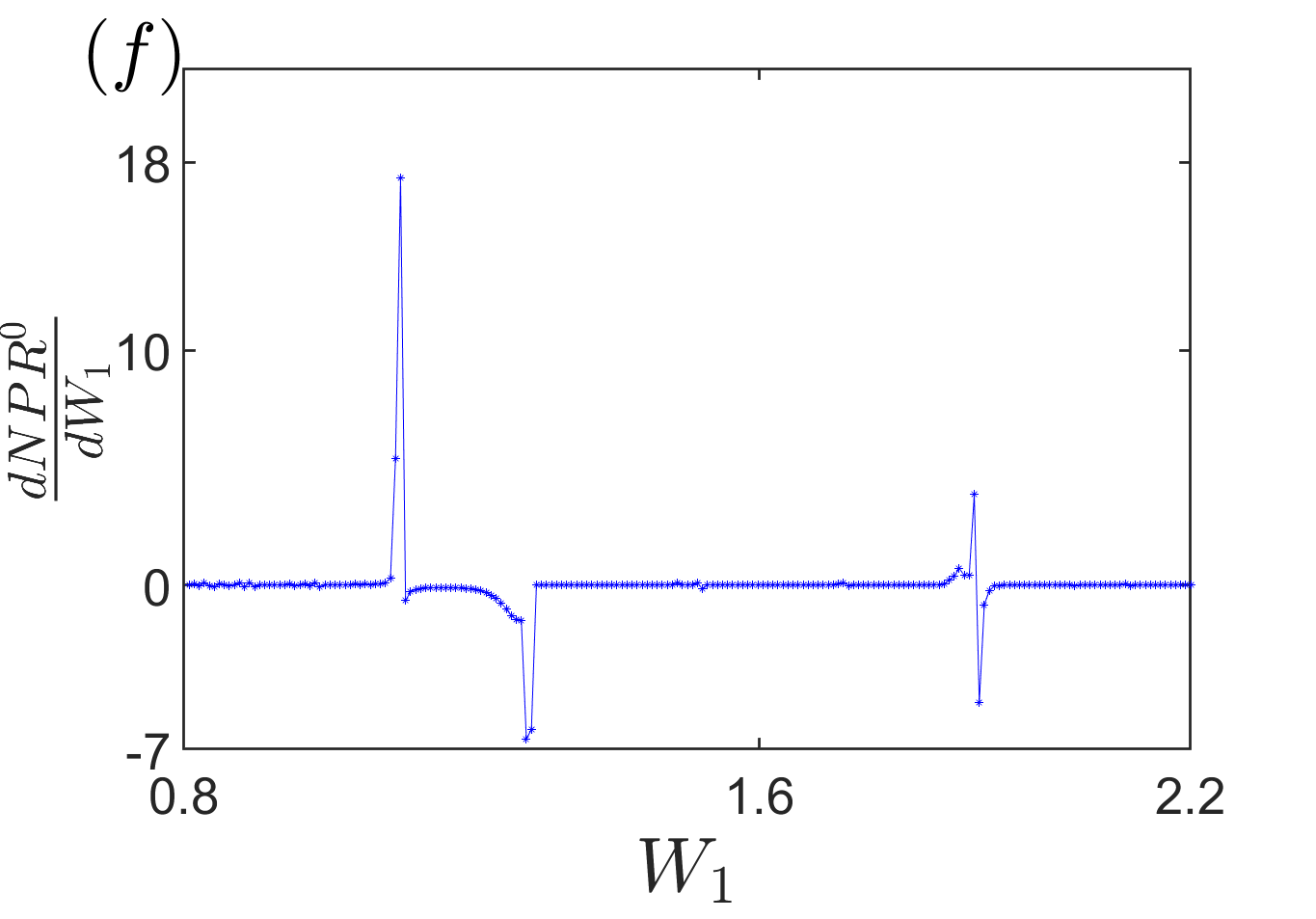}
\caption{(Color online) (a) The topological invariant versus the parameter $W_{1}$. (b) The $IPR^{B}$ (red curve) and $NPR^{B}$ (blue curve) being plotted as a function of $W_{1}$. The shaded region indicates the coexistence regime. (c) The real part of bulk energy spectrum, where the different colors stand for the different values of {$IPR^{|\Psi^{n}\rangle}$}. (d) The evolution of the topological zero energy edge modes $|E|^{0}$ via the $W_{1}$, associated with the $IPR^{0}$ and $NPR^{0}$. (e) The detailed information of the localization of the eigenstates with the lowest energy. (f) the derivative of $NPR^{0}$, where there exists some discontinuity points. From (a) to (d) $L=2000$ and $l=0.2L$, $L=10000$ for (e) and (f). The same parameters are $t_{1}=1$, $t_{2}=1.3$, $\gamma=0.05$ and $W_{2}=W_{1}$.}
\label{fig1}
\end{figure}

In order to be accurate of exploring the localization properties of the bulk and edge state, the trivial and nontrivial winding number of the system can be shown in Fig. \ref{fig1}(a) firstly. In the following, we begin with the bulk eigenstates.

$The$ $Bulk$. Fig. \ref{fig1}(b) exhibits $IPR^{B}$ (red curve) and $NPR^{B}$ (blue curve) for $L=2000$. $IPR^{B}$ and $NPR^{B}$ are clearly simultaneously nonzero only in one region $0.2<W_{1}<1.28$, which implies that the localized and extended regions coexist in the bulk. Similar to Refs. [\onlinecite{Li50}, \onlinecite{Xiao51}], this regime is marked as $ME^{B}$. Further, the bulk energy spectra encoded with $IPR^{|\Psi^{n}\rangle}$ are exhibited in Fig. \ref{fig1}(c), in which the values of $IPR^{|\Psi^{n}\rangle}$ are degrees of localization of the associated bulk eigenstate. We first focus on the clean limit $W_{1}=0$, where the system degenerates into the SSH model with a nontrivial phase ($\mu=1$ shown in Fig. \ref{fig1}(a)) and the bulk eigenstates are extended averagely at this point [\onlinecite{As3}]. The bulk eigenstates remain extended up to the limit of quasidisorder at $W_{1}=0.2$. As $W_{1}$ increases from $0.2$ to $1.28$, an increasing number of extended eigenstates tend to transform into localized ones. That is, the extended eigenstates can coexist with the localized ones in the bulk. The dark blue color of the energy spectrum vanishes eventually when $W_{1}>1.28$, which states that all bulk eigenstates are localized forever in large $L$ limit.

$The$ $Edge$. To seek the localization behaviors of the edge modes explicitly, $IPR^{0}$ and $NPR^{0}$ are shown as a function of $W_{1}$ in Figs. \ref{fig1}(d) and (e). For $W_{1}=0$, $IPR^{0}\neq0$ and $NPR^{0}=0$ means that the associated state is localized [\onlinecite{As3}]. When the quasidisorder strength is in the range $0<W_{1}<1.09$, $IPR^{0}$ decreases in the form of high-frequency oscillations and $NPR^{0}$ remains zero with increasing disorder strength. Interestingly, as shown in Fig. \ref{fig1}(e), the decrease of $IPR^{0}$ to zero is simultaneous with the increase of $NPR^{0}$ to nonzero values at $W_{1}=1.09$, which declares the state itself undergoes a sharp localization transition. Meanwhile, the topological zero-energy mode vanishes at the point $W_{1}=1.09$, unveiling a topological phase transition. Unexpectedly, in the trivial regime of $1.09<W_{1}<1.905$, the eigenstate ever experiences a localized transition from $IPR^{0}=0$ and $NPR^{0}\neq0$ to $IPR^{0}\neq0$ and $NPR^{0}=0$ near $W_{1}=1.276$, i.e., from the extended to localized ones. In addition, when the quasidisorder strength goes around $W_{1}=1.905$, $NPR^{0}$ changes as zero-nonzero-zero, and $IPR^{0}$ changes from nonzero-zero-nonzero at the same time, which indicates a localized transition at this point. Moreover, the lowest branch of the absolute energy becomes zero again, which implies a topological phase transition at $W_{1}=1.905$.

As shown in Fig. \ref{fig1}(e), $NPR^{0}$ changes from zero to nonzero at the localized transition point, or vice versa. Note that $\max[NPR^{0}]\approx0.17$ with $W_{1}\in[0,4]$, which means that the change in the magnitude of $NPR^{0}$ is very small. To characterize the localization points more explicitly, the derivative of $NPR^{0}$ with respect to $W_{1}$ is presented numerically in Fig. 1(f), it can be easily found that $\frac{dNPR^{0}}{dW_{1}}$ becomes very sharp and looks like a $\delta$ function around four points. Compared with Figs. \ref{fig1}(d) and (e), a localization phase transition can be considered to take place at these points.

\section{Reentrant Localized Bulk}\label{IV}

\begin{figure}
\includegraphics[width=4.2cm,height=3.8cm]{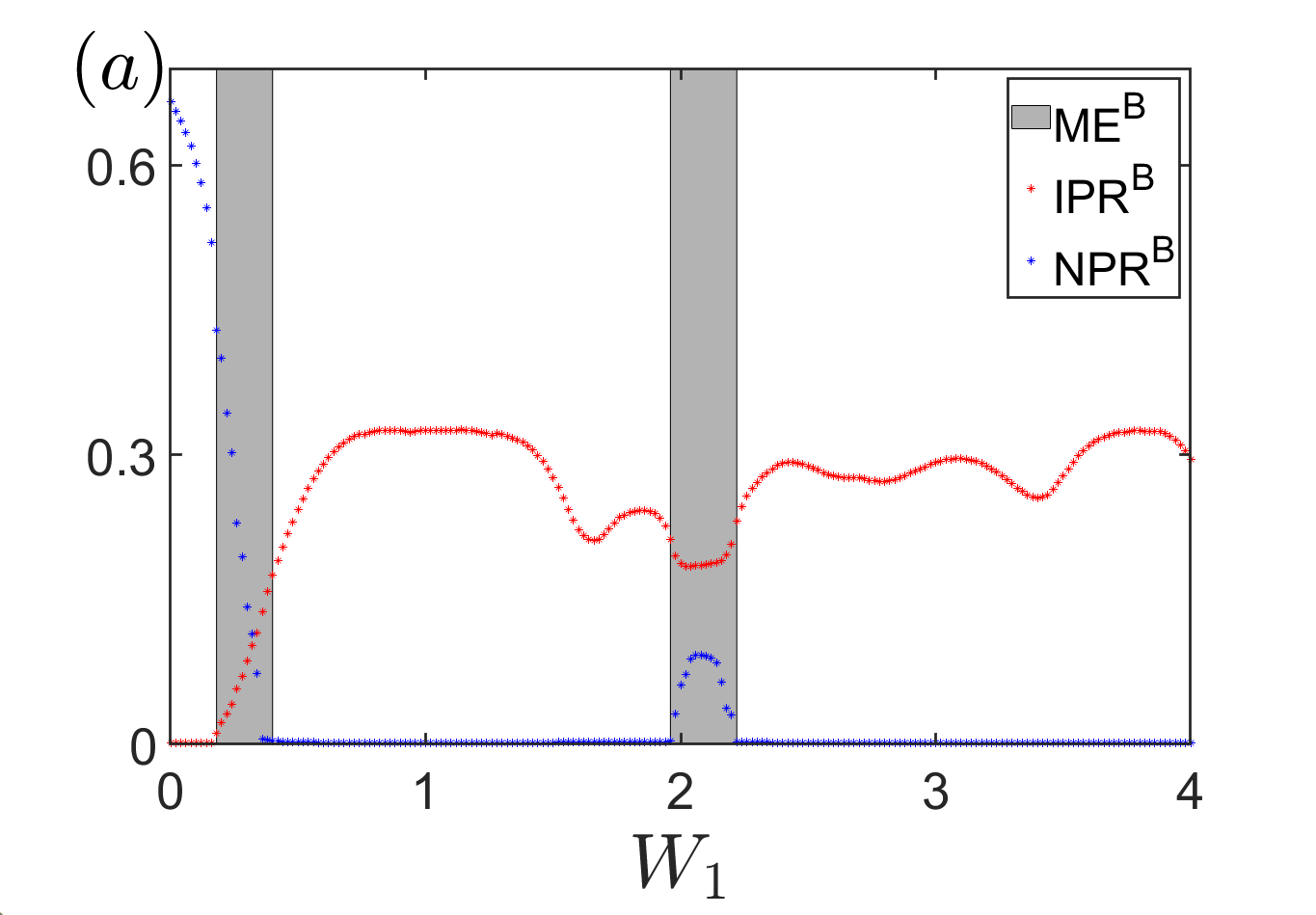}
\includegraphics[width=4.2cm,height=3.8cm]{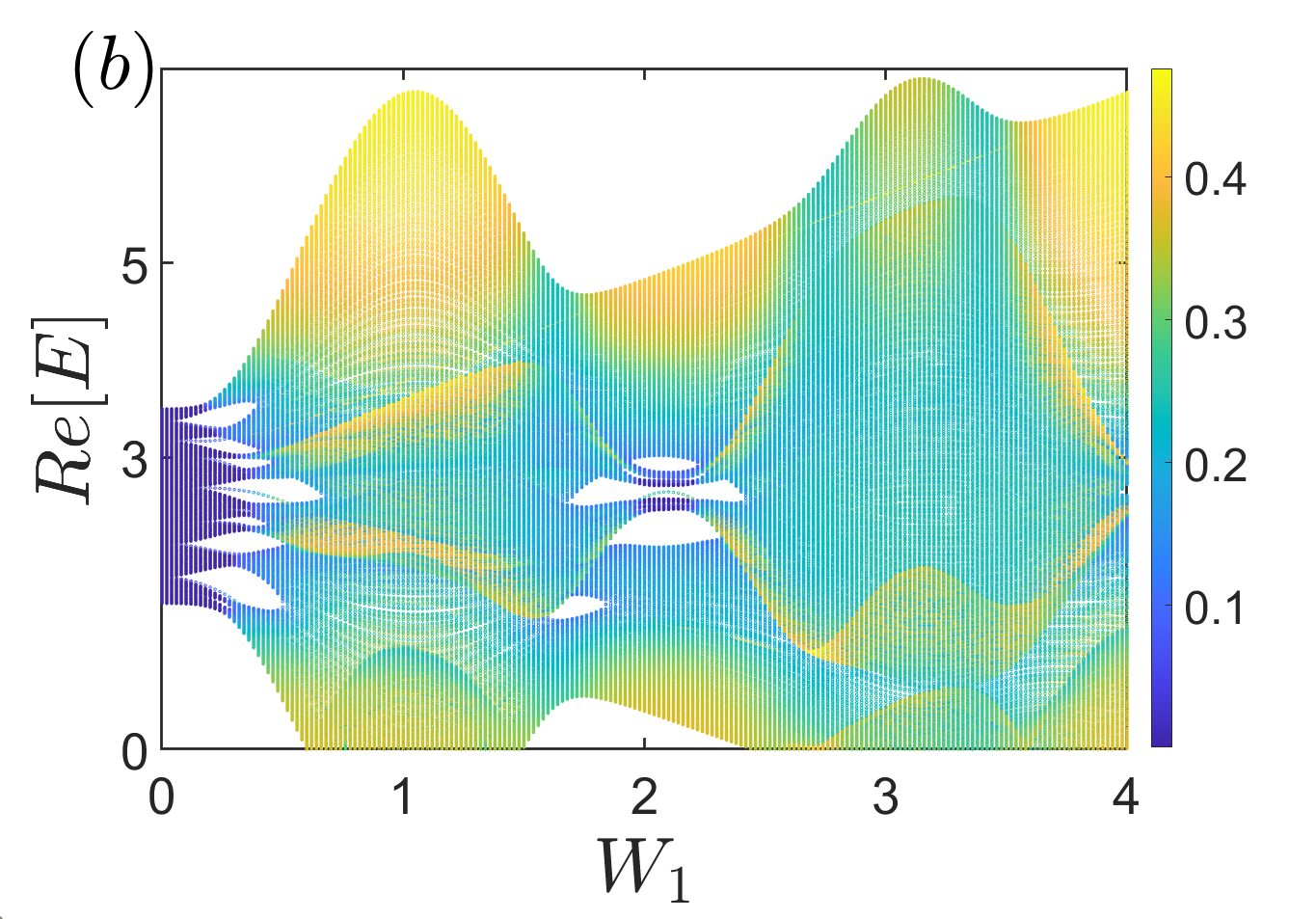}

\includegraphics[width=4.2cm,height=3.8cm]{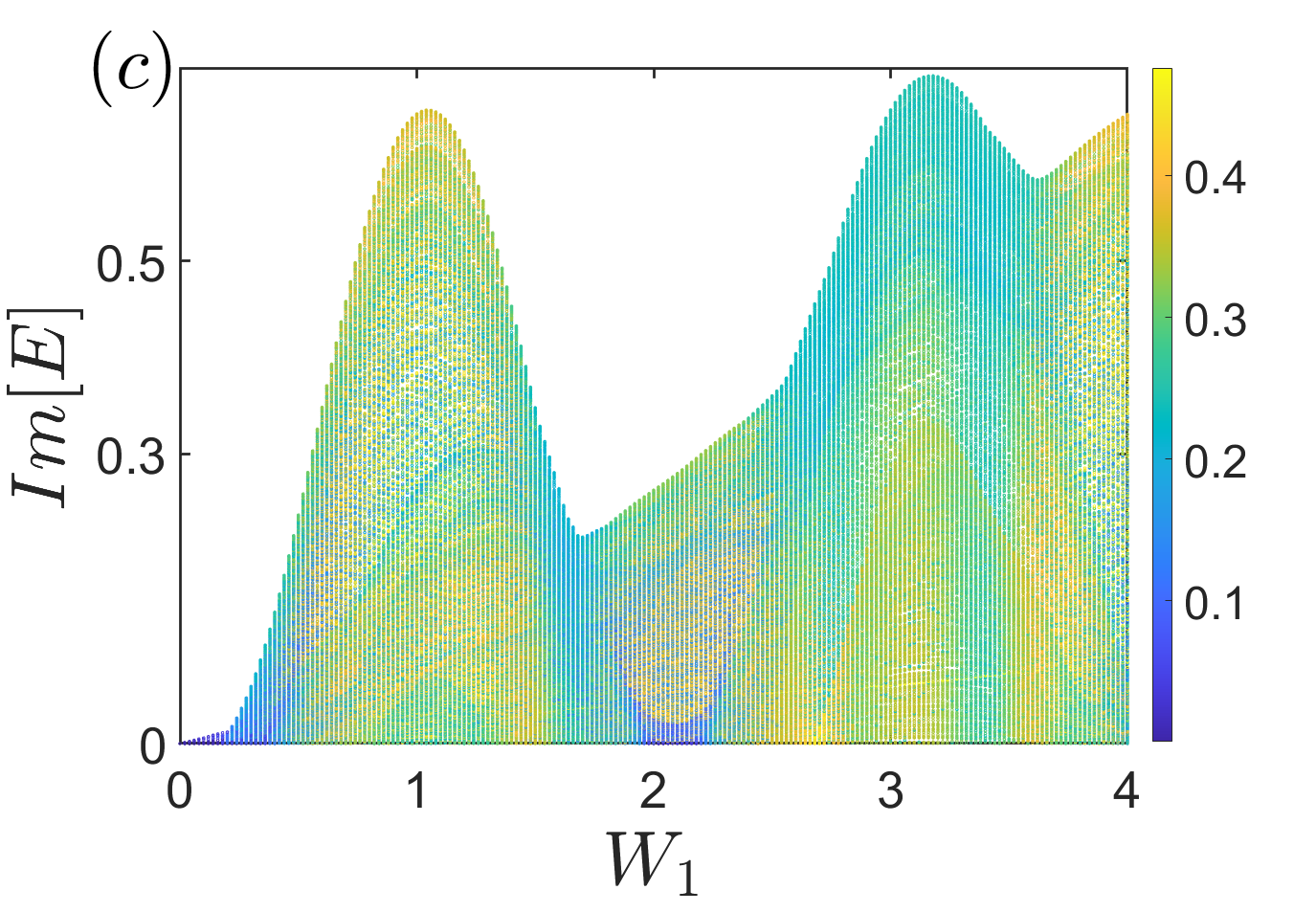}
\includegraphics[width=4.2cm,height=3.8cm]{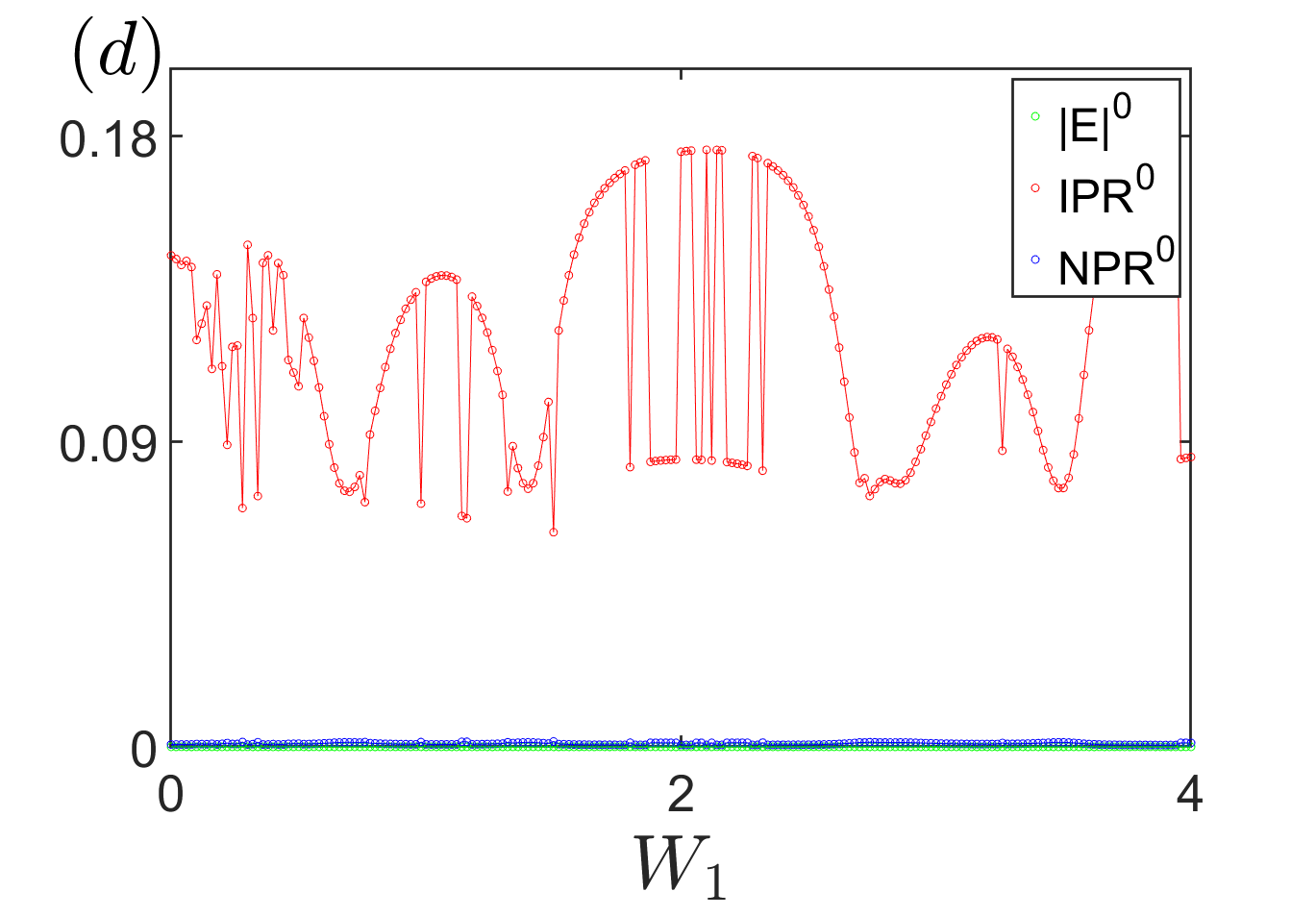}
\caption{(Color online) (a) $IPR^{B}$ (red curve) and $NPR^{B}$ (blue curve) versus $W_{1}$. Shaded regions stand for the intermediate regimes. (b) and (c) The real and imaginary parts of bulk energy spectrum, where the dressed colors stand for the different values of {$IPR^{|\Psi^{n}\rangle}$} for each bulk eigenstates. (d) The topological edge modes $|E|^{0}$, accompanied by the corresponding $IPR^{0}$ and $NPR^{0}$. The common parameters are $t_{1}=1$, $t_{2}=2.5$, $\gamma=0.2$, $L=2000$ and $W_{2}=-2\cos(3W_{1})+2$.}
\label{fig2}
\end{figure}

In Sec. \ref{III}, we have shown that there is only one region where the localized and extended eigenstates coexist in the bulk. Surprisingly, by together analyzing $IPR^{B}$ and $NPR^{B}$ in Fig. \ref{fig2}(a), we discern that the reentrant localization may be realized under the modulation of the quasi-periodic controlling parameters. The fundamental changes of the localization nature also can be reflected by system energies. In Fig. \ref{fig2}(b), the real part of bulk energy spectra and the corresponding $IPR^{|\Psi^{n}\rangle}$ are plotted. For $0<W_{1}<0.18$, the bulk eigenstates spread over the chain i.e., an extended eigenstate appears. For $0.18<W_{1}<0.38$, the mapped color for the bulk eigenstates will change gradually from dark blue to cyan, which means that an increasing number of extended eigenstates change into localized ones. When $W_{1}=0.38$, the bulk eigenstates fully become localized. Unexpectedly, some of the localized bulk eigenstates begin to revert to extended ones in the range of $1.96<W_{1}<2.22$, and the bulk is in the coexistence of localized and extended eigenstates again. As $W_{1}$ increases further, all bulk eigenstates become localized for $W_{1}>2.22$. Note that the proportion of reentrant extended eigenstates is small compared to the total bulk eigenstates, which also can be mirrored by the imaginary part of the bulk energy spectrum, as shown in Fig. \ref{fig2}(c). As $W_{1}$ increases over the regime $0\leq W_{1}<0.18$, the imaginary part of the energies increases slowly, and the corresponding dark-blue-colored eigenstates are extended. When $W_{1}$ goes into the interval $(0.18, 0.38)$, the mixed colors clarify that the bulk eigenstates of different degrees of localization are intermingled with each other. Remarkably, if the quasidisorder $W_{1}$ is set in the regime $(1.96, 2.22)$, we find that the extended eigenstates exist almost in the lowest imaginary energies, which matches well with the results shown in Figs. \ref{fig2}(a) and \ref{fig2}(b). The localization of edge states also can be explored in Fig. \ref{fig2}(d), where the edge modes always exist over the entire range of parameters. On the other hand, $IPR^{0}\neq0$ and $NPR^{0}=0$ are always satisfied, which implies the corresponding state is localized as well. The oscillatory behavior of $IPR^{0}$ indicates that even though every edge state is assembled at the boundary, the degrees of localization are quite different and are not positively correlated with the disorder strength.

\begin{figure*}[!htbp]
\includegraphics[width=4.2cm,height=3.8cm]{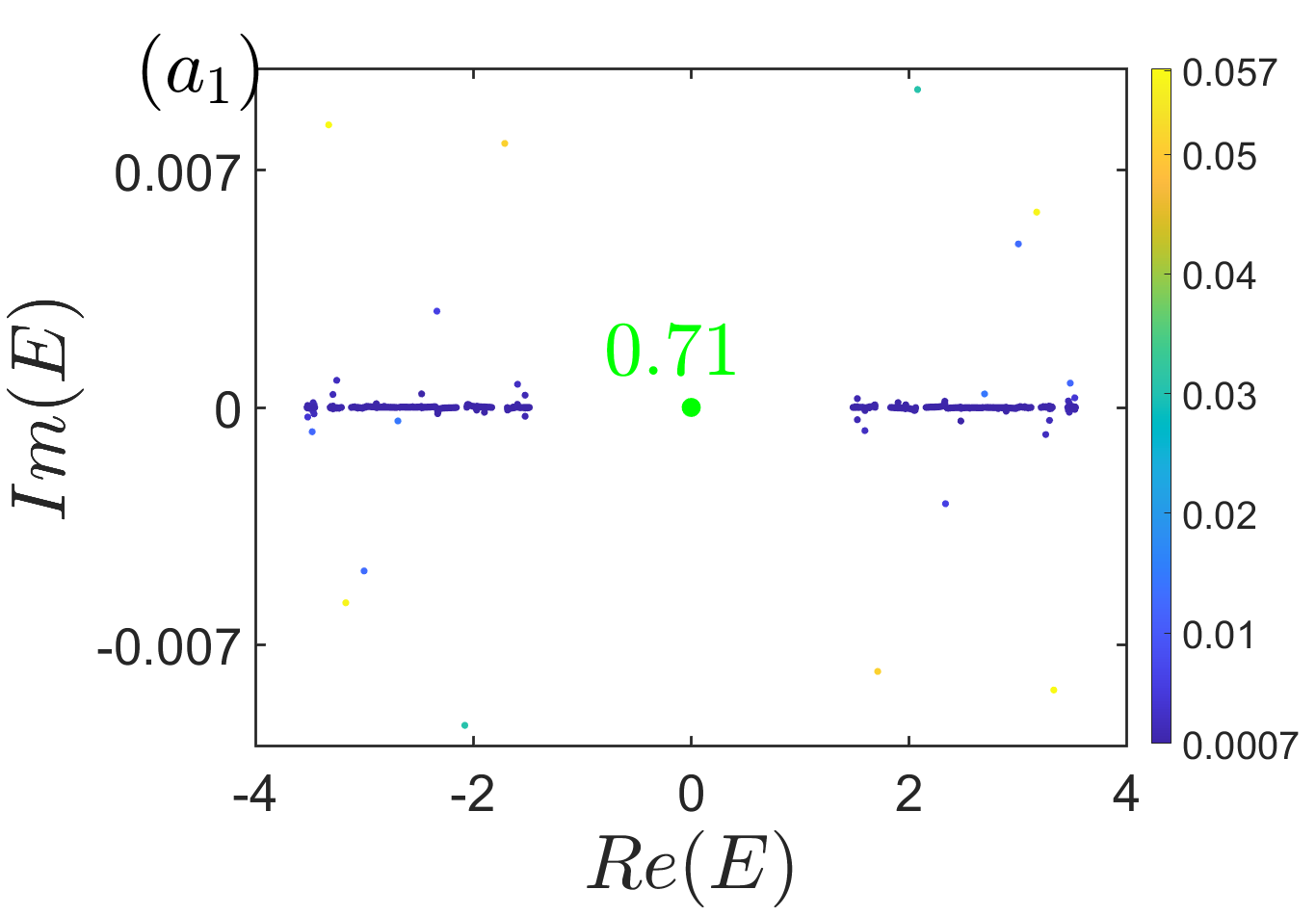}
\includegraphics[width=4.2cm,height=3.8cm]{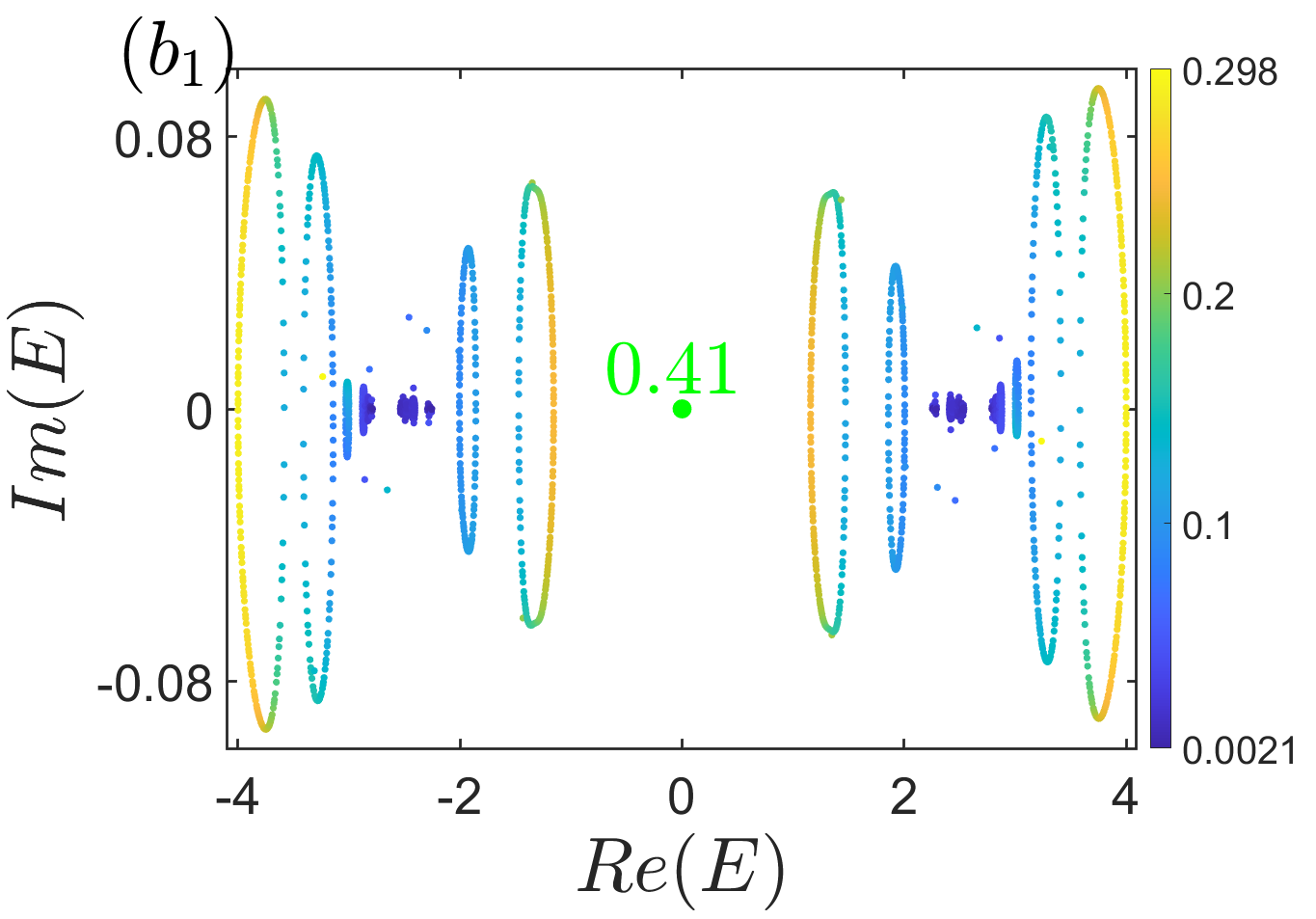}
\includegraphics[width=4.2cm,height=3.8cm]{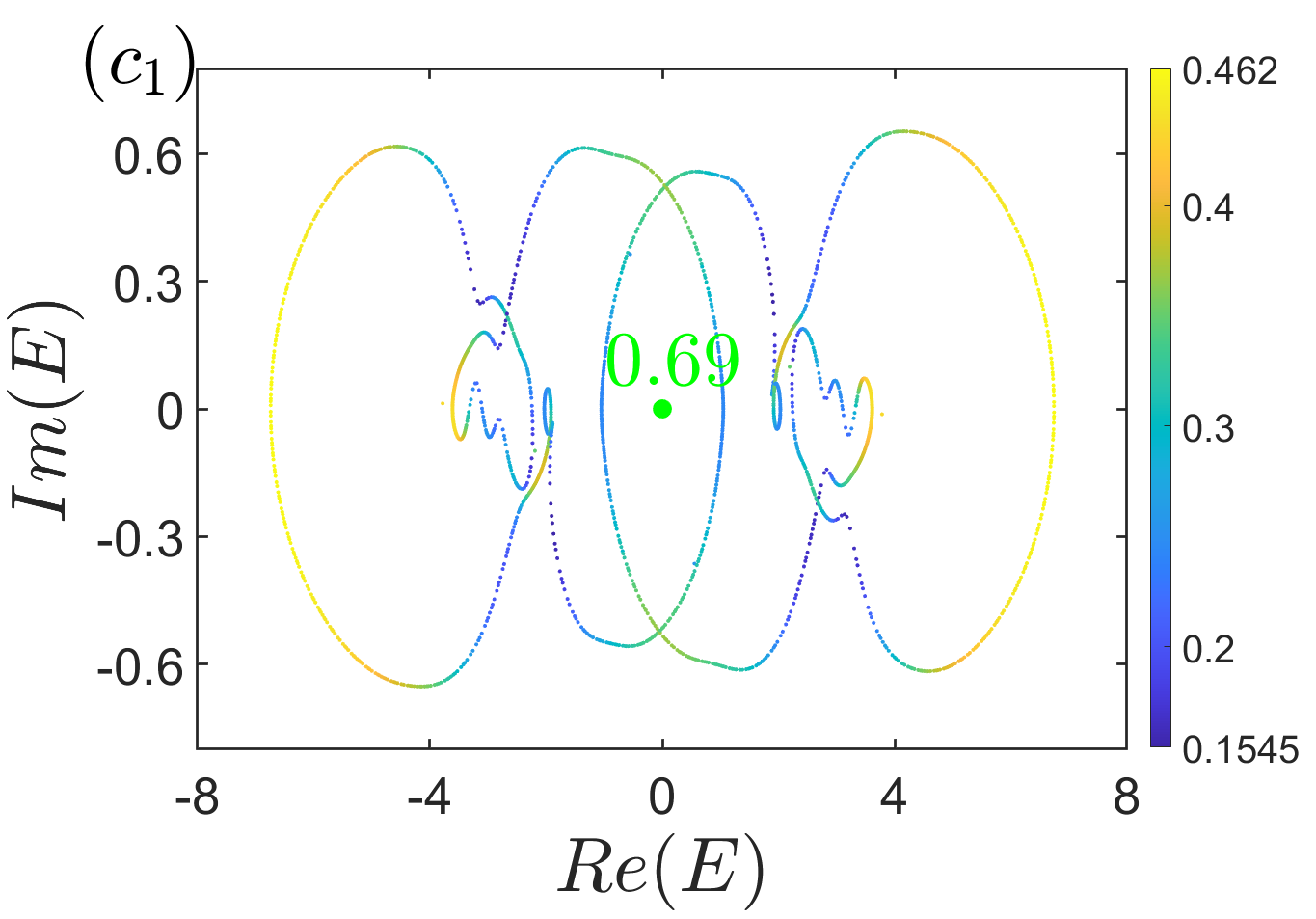}
\includegraphics[width=4.2cm,height=3.8cm]{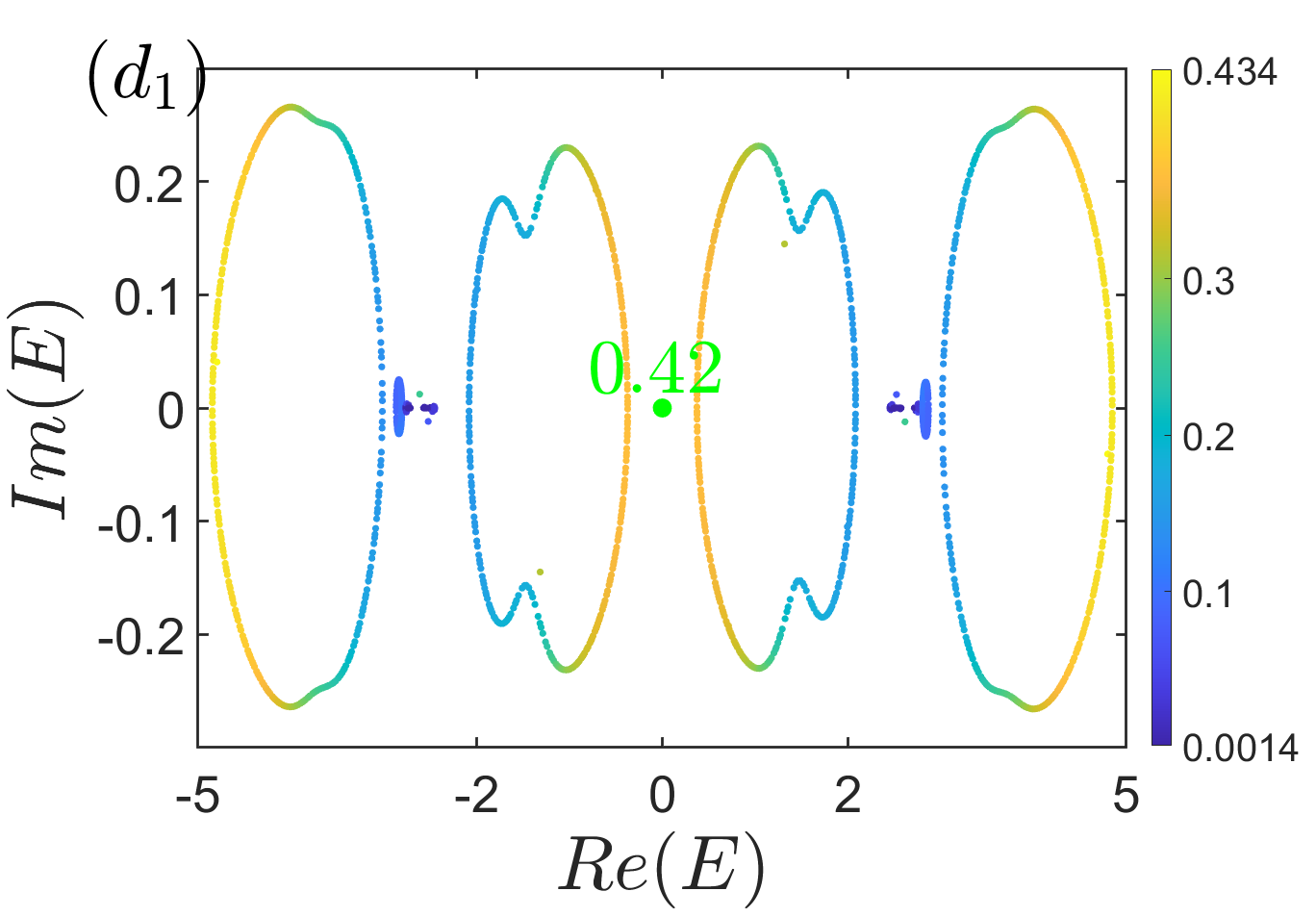}

\includegraphics[width=4.2cm,height=3.8cm]{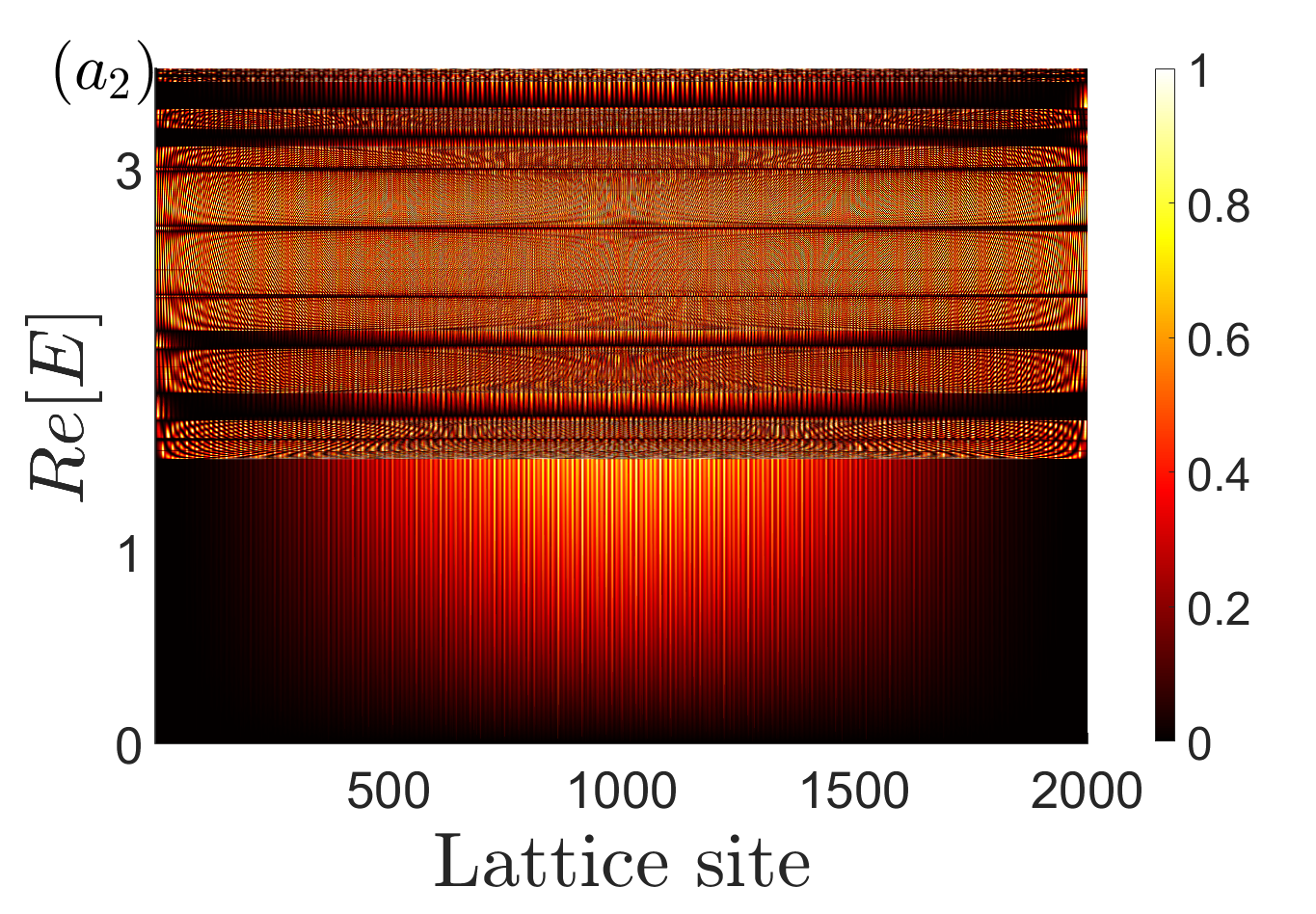}
\includegraphics[width=4.2cm,height=3.8cm]{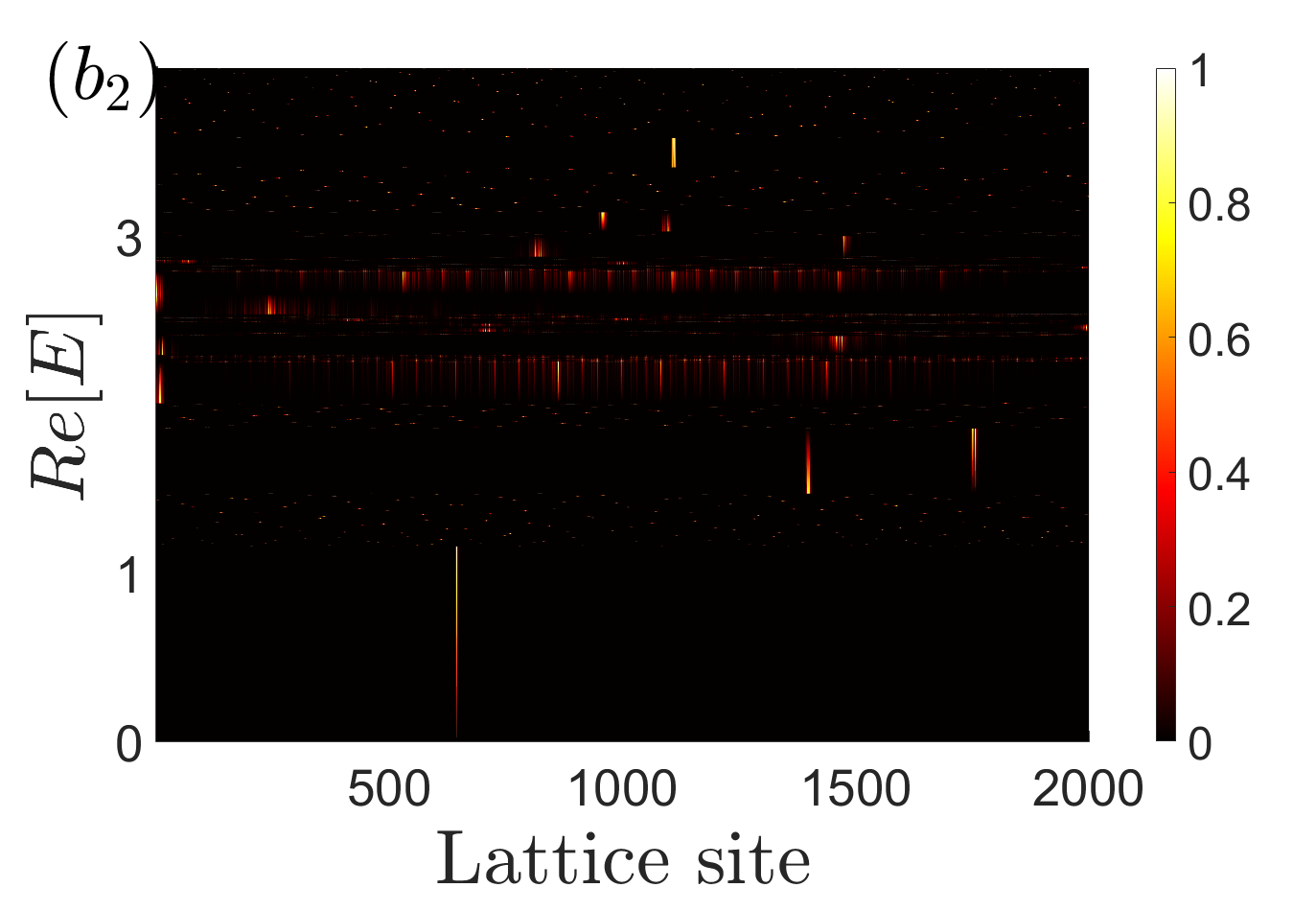}
\includegraphics[width=4.2cm,height=3.8cm]{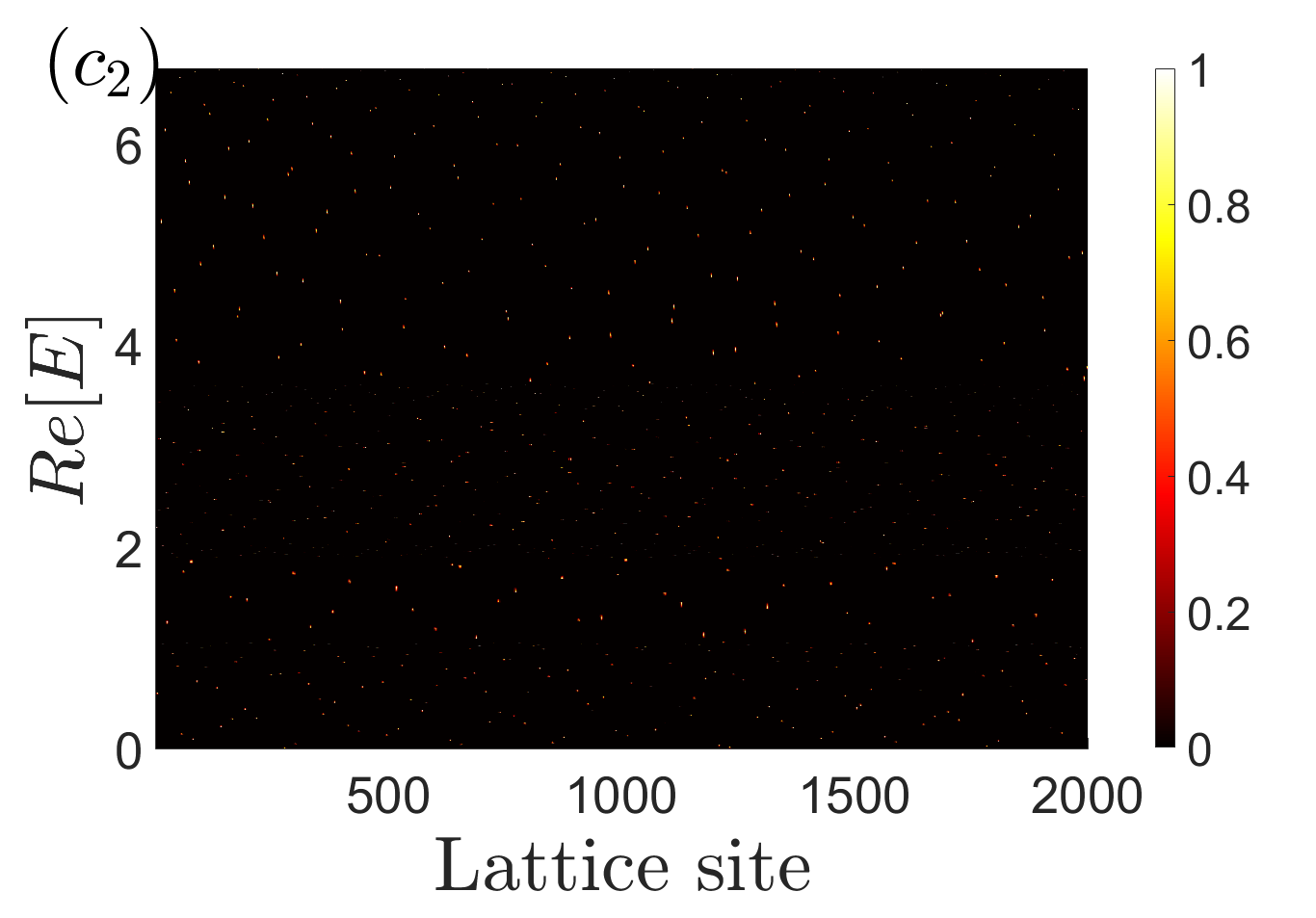}
\includegraphics[width=4.2cm,height=3.8cm]{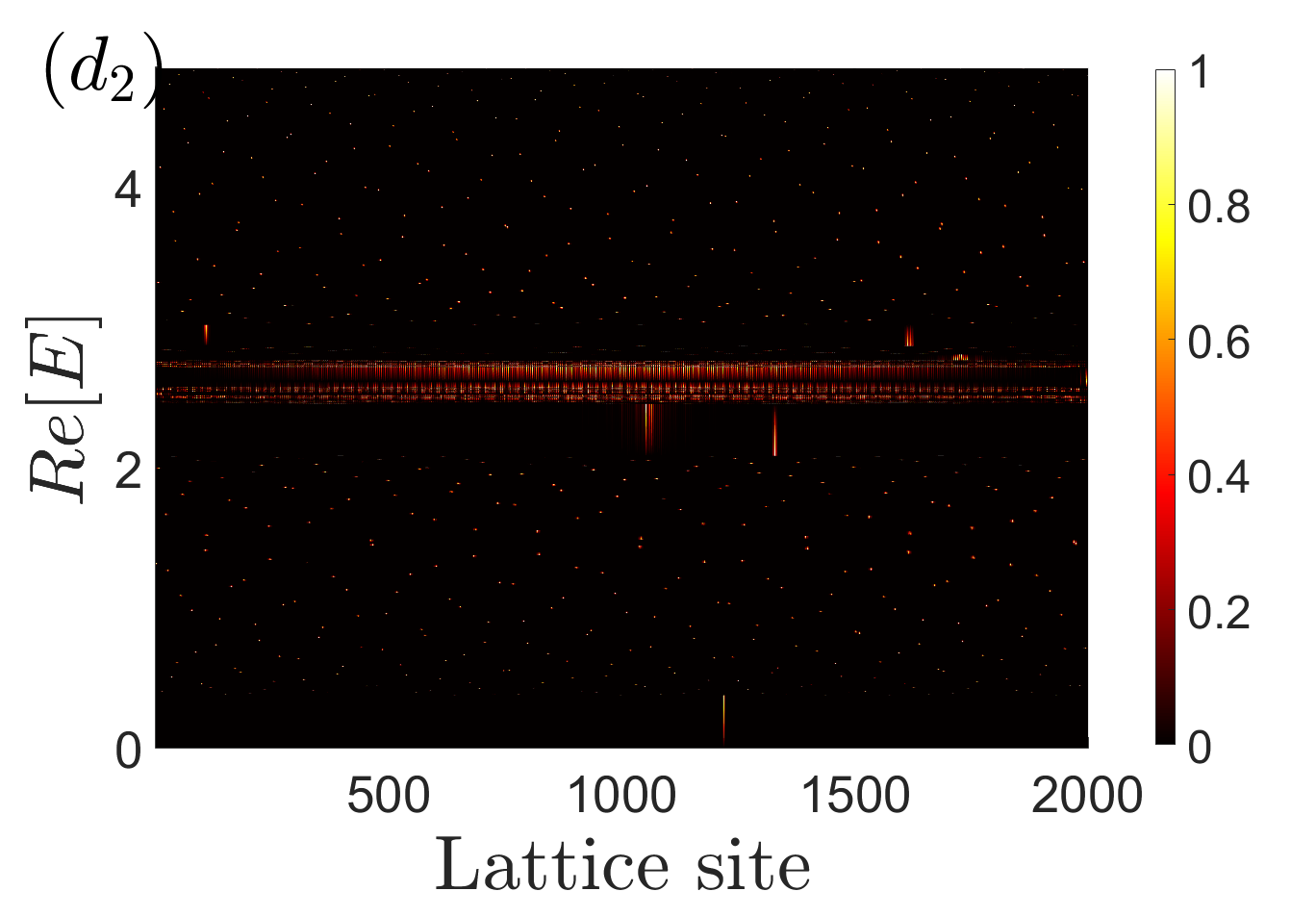}

\includegraphics[width=4.2cm,height=3.8cm]{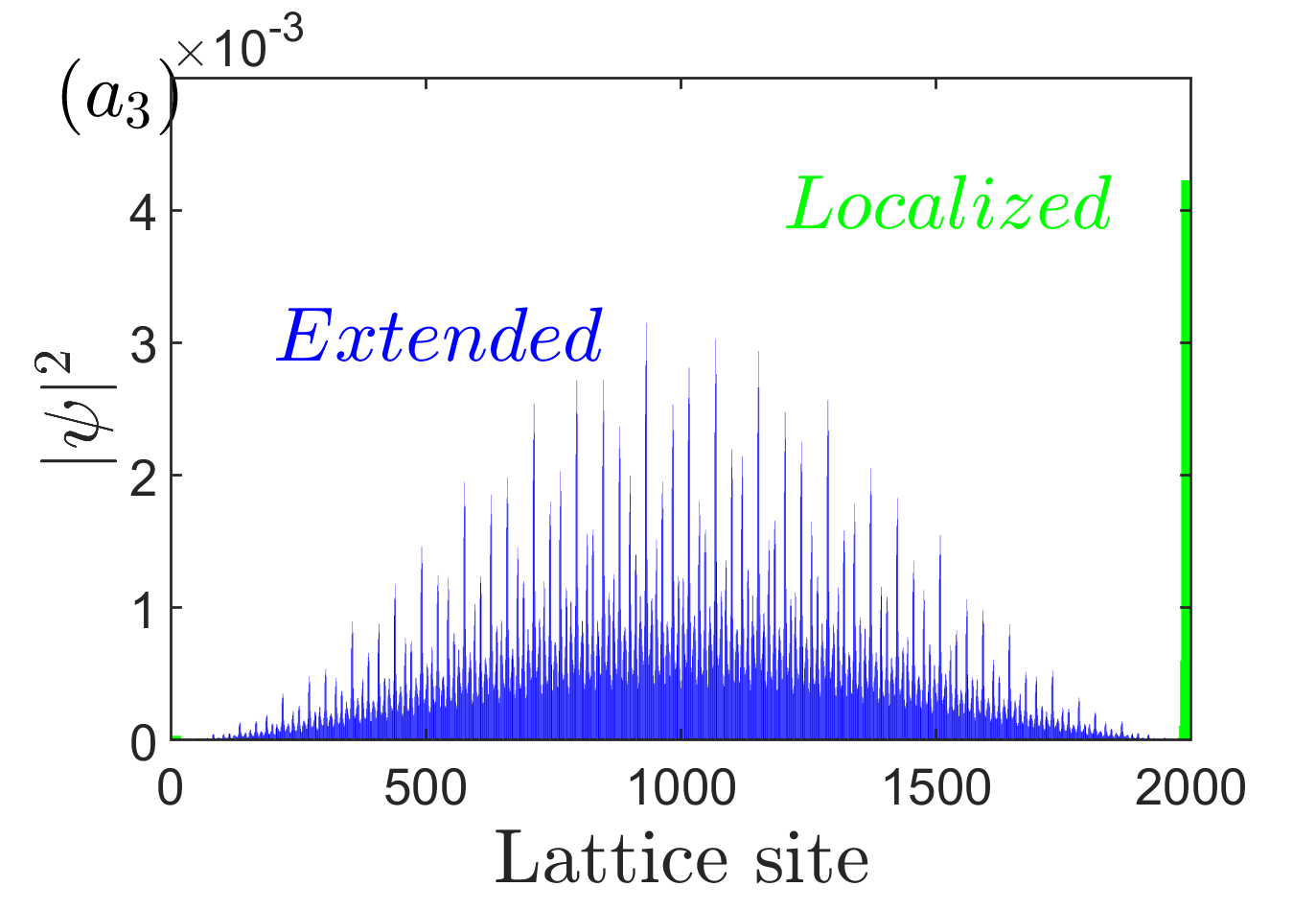}
\includegraphics[width=4.2cm,height=3.8cm]{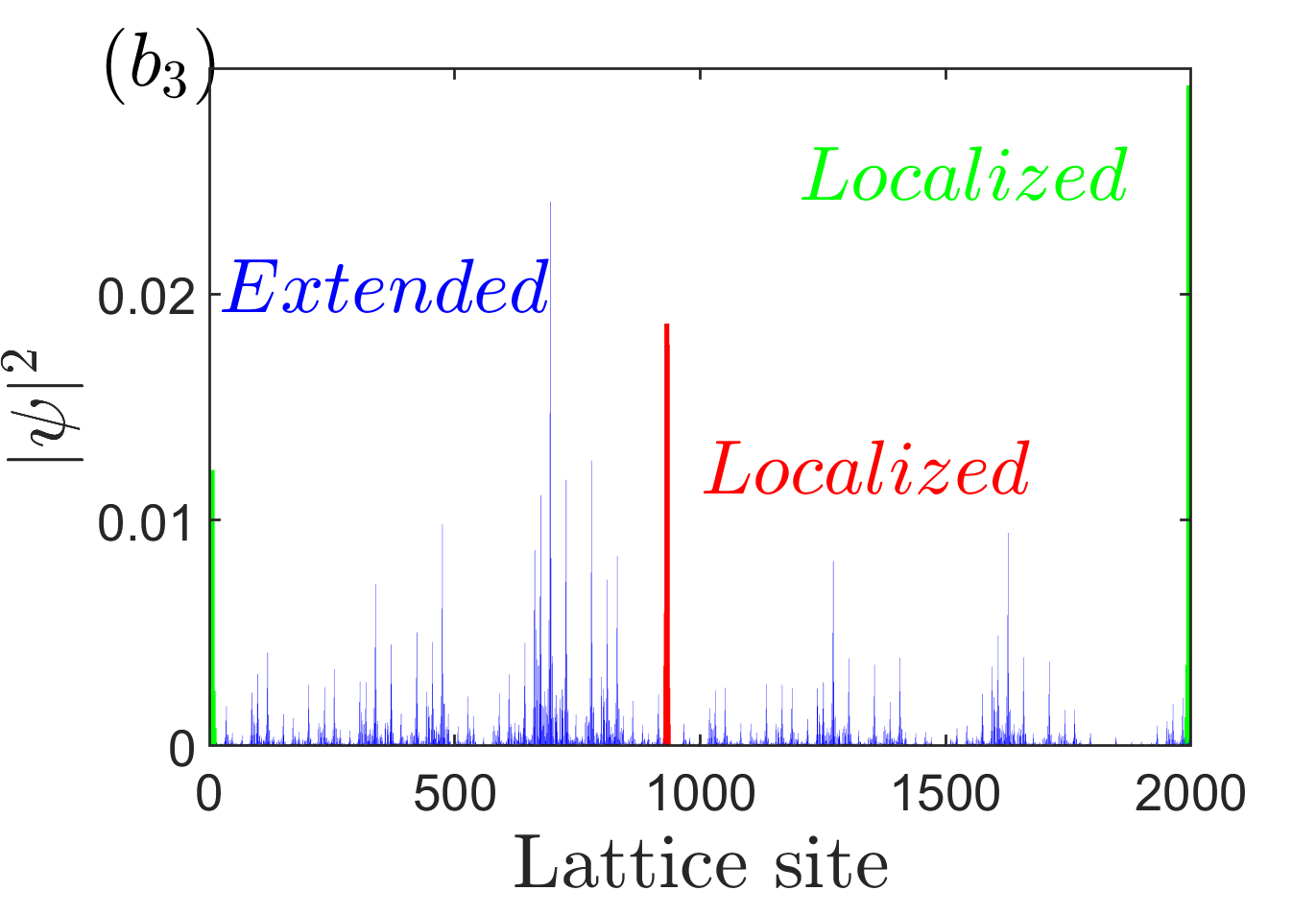}
\includegraphics[width=4.2cm,height=3.8cm]{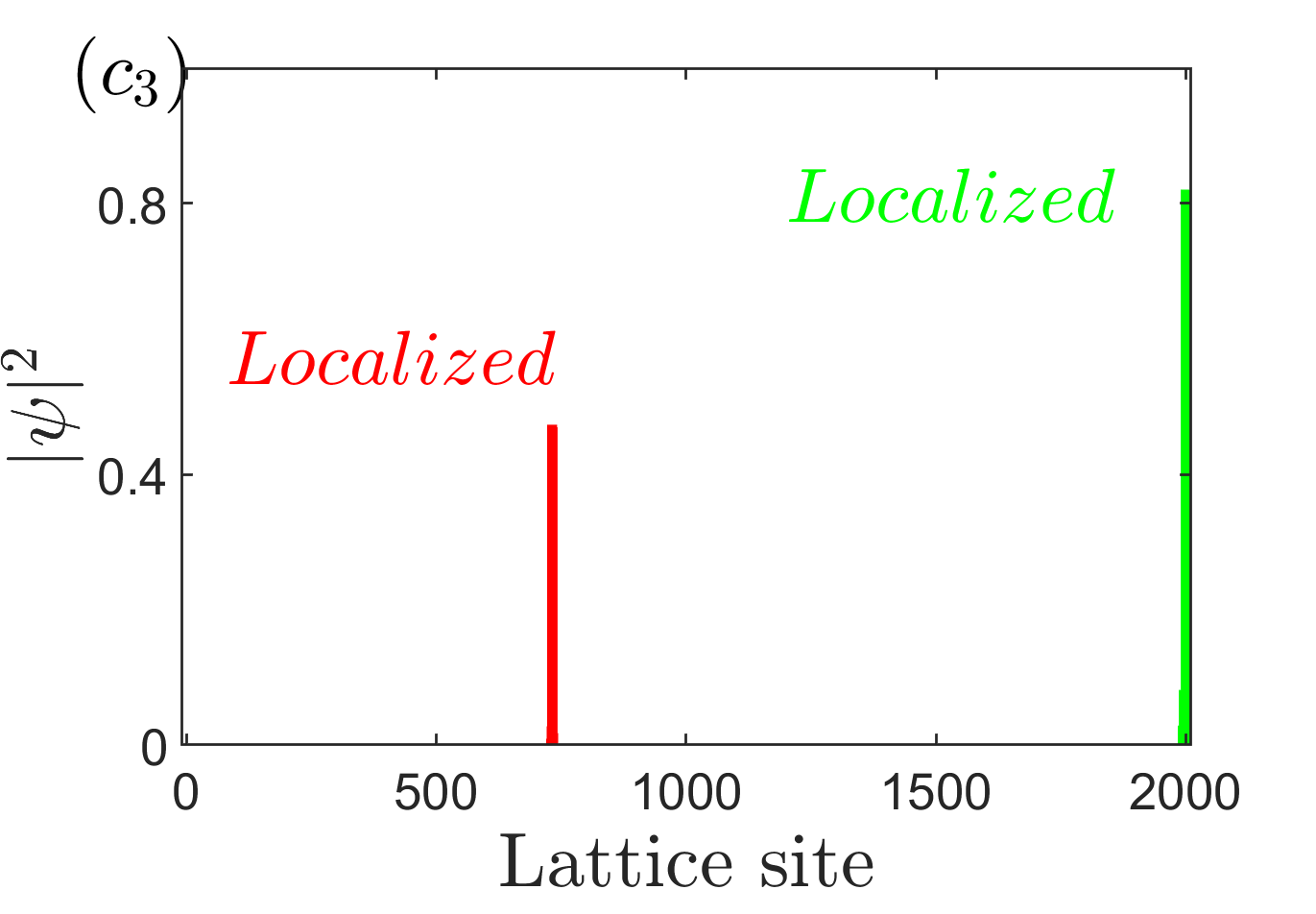}
\includegraphics[width=4.2cm,height=3.8cm]{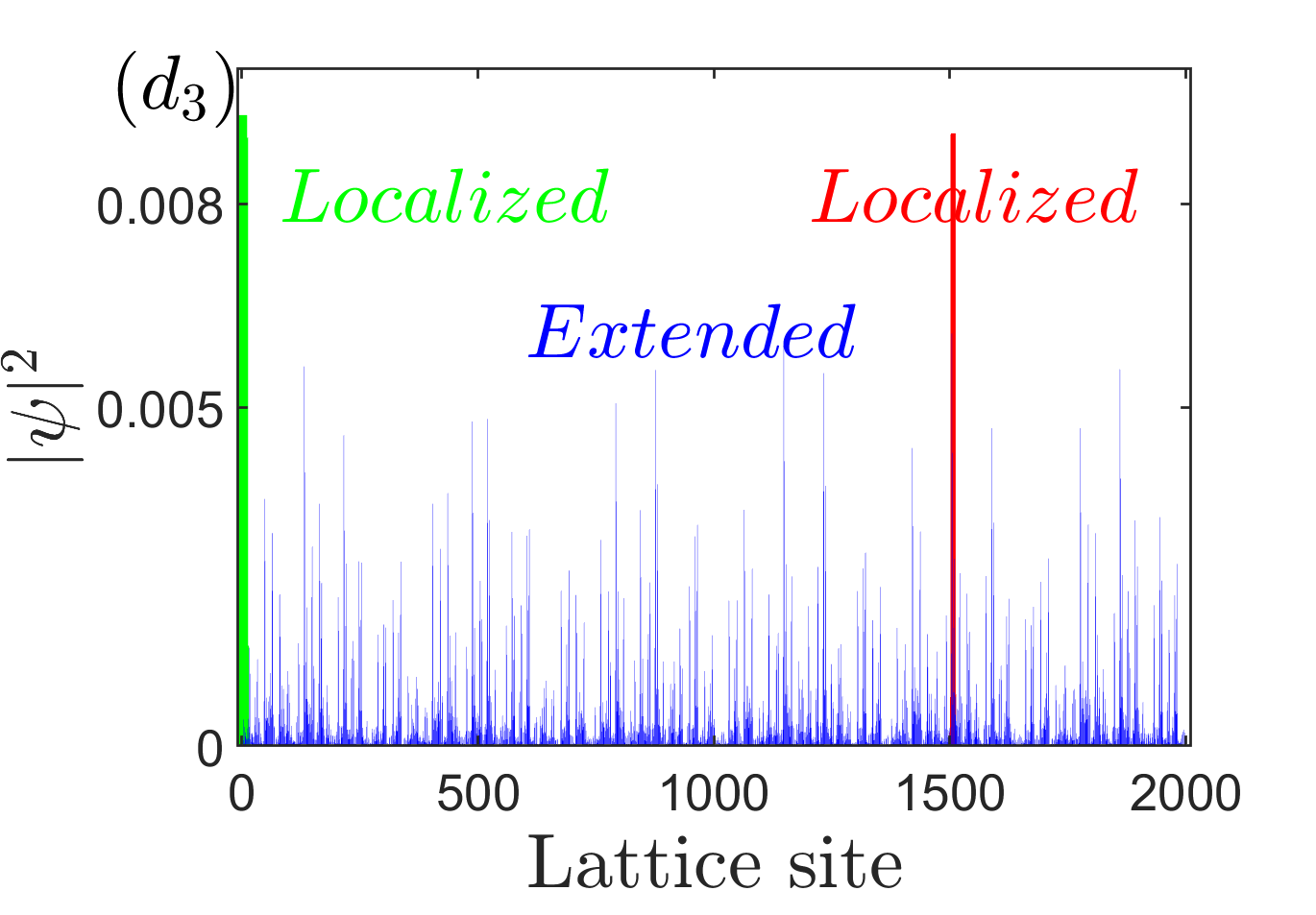}
\caption{(Color online) (a$_{1}$-d$_{1}$) The energy spectrum on the complex plane with the corresponding {$IPR^{|\Psi^{n}\rangle}$}. (a$_{2}$-d$_{2}$) The normalized bulk eigenstates with the change of eigenenergies. (a$_{3}$-d$_{3}$) Both the blue and red lines stand for the localized properties of some bulk eigenstates, and the zero-energy state is shown as the green bar. For a better visibility, the values have been scaled proportionally. For (a$_{1}$-a$_{3}$), $W_{1}=0.15$, all bulk eigenstates are extended. For (b$_{1}$-b$_{3}$), $W_{1}=0.35$, some eigenstates are still extended while some are localized. For (c$_{1}$-c$_{3}$), $W_{1}=1$, all bulk eigenstates are transformed into localized. For (d$_{1}$-d$_{3}$), $W_{1}=2.02$, some already localized states will change into extended. Other parameters are same as the ones in Fig. \ref{fig2}.}
\label{fig3}
\end{figure*}

\begin{figure}
\includegraphics[width=4.2cm,height=3.8cm]{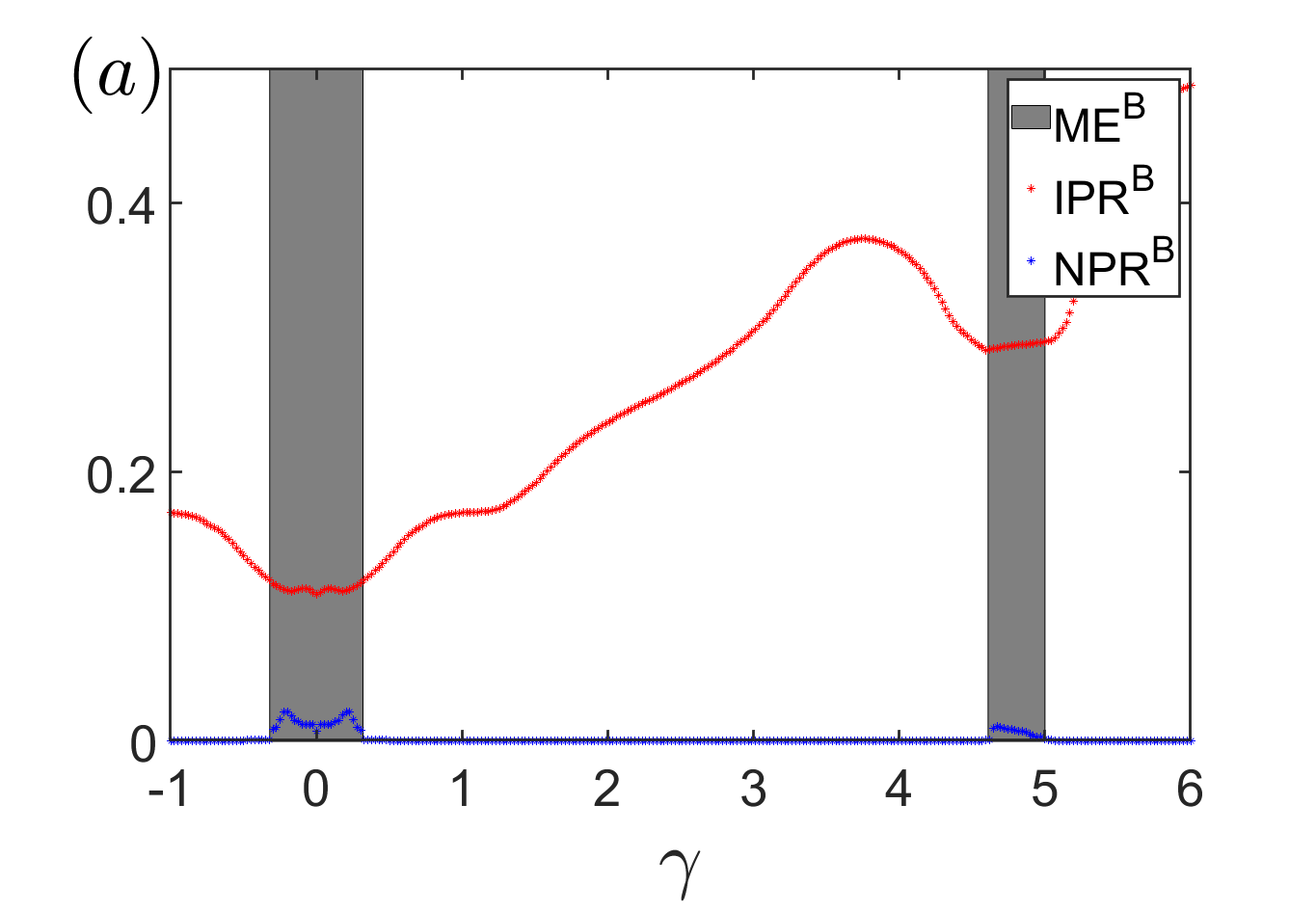}
\includegraphics[width=4.2cm,height=3.8cm]{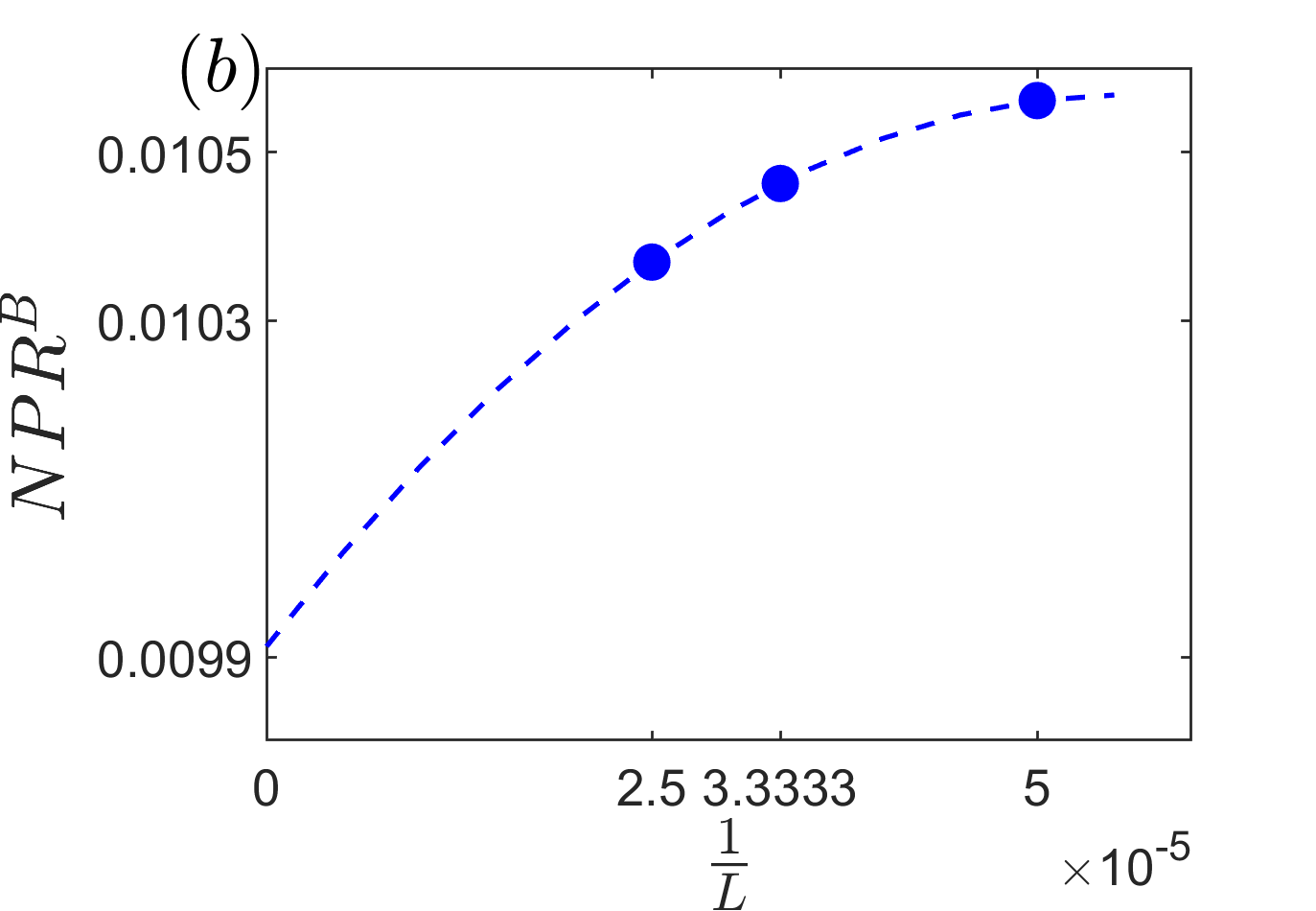}

\includegraphics[width=4.2cm,height=3.8cm]{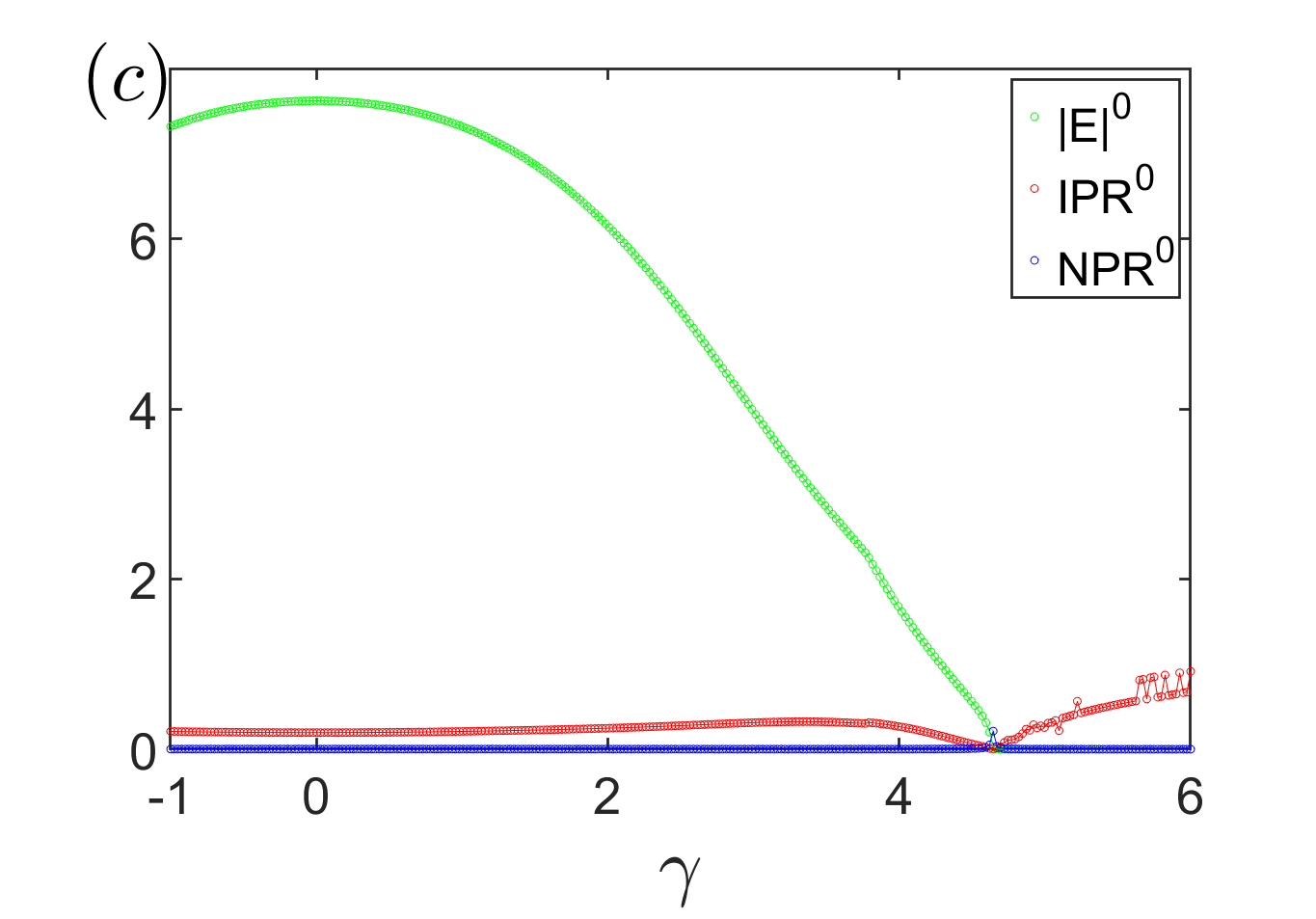}
\includegraphics[width=4.2cm,height=3.8cm]{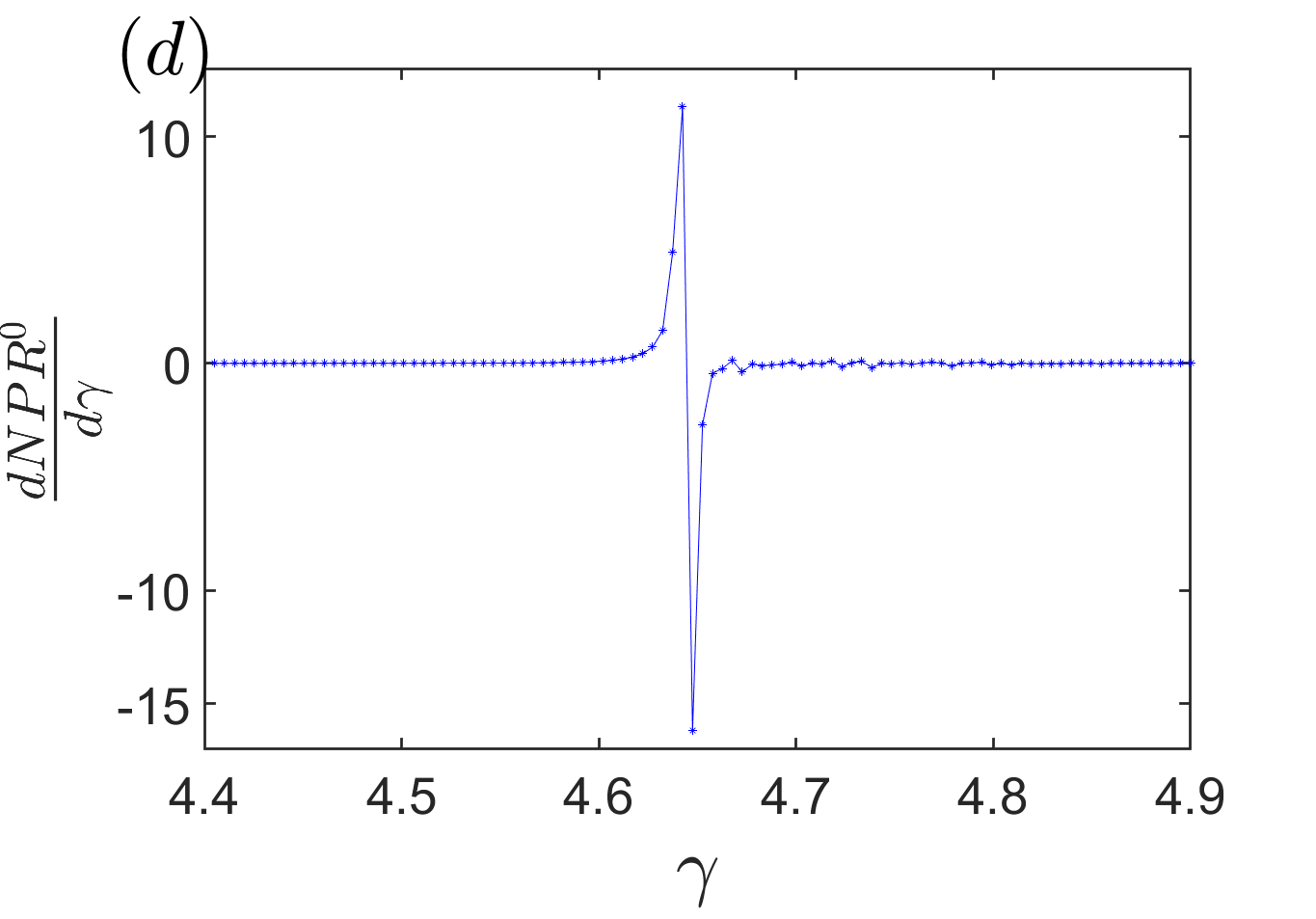}
\caption{(Color online) (a) $L=12000$, $IPR^{B}$ (red curve) and $NPR^{B}$ (blue curve) evolved with $\gamma$. The two shaded regions stand for the intermediate regimes. (b) The finite size scaling analysis of $NPR^{B}$ for $\gamma=4.66946$. $L=20000$, $30000$ and $40000$, respectively. (c) The evolution of the topological zero energy edge modes $|E|^{0}$ versus the $\gamma$, associated with the $IPR^{0}$ and $NPR^{0}$, $L=2000$. (d) The derivative of $NPR^{0}$ by the change of $\gamma$ with $L=12000$. The common parameters are $t_{1}=9$, $t_{2}=1$, $W_{1}=0.0039$ and $W_{2}=1.563$.}
\label{fig4}
\end{figure}

Concretely, the localization behaviors of the bulk and edge can be demonstrated at different statuses in Fig. \ref{fig3}. In Figs. \ref{fig3}(a$_{1}$)-\ref{fig3}(a$_{3}$), the bulk is extended with $W_{1}=0.15$. For a large quasidisorder $W_{1}=0.35$, Figs. \ref{fig3}(b$_{1}$)-\ref{fig3}(b$_{3}$) exhibit that the coexistence occurs. Further, Figs. \ref{fig3}(c$_{1}$)-\ref{fig3}(c$_{3}$) show that the bulk becomes localized at $W_{1}=1$, and Figs. \ref{fig3}(d$_{1}$)-\ref{fig3}(d$_{3}$) shows that coexistence finally occurs in the bulk again at $W_{1}=2.02$. The red dot in Figs. \ref{fig3}(a$_{1}$-d$_{1}$) stands for the zero-energy modes $|E|^{0}$, and the associated number is the value of the $IPR^{0}$. The bulk eigenenergies along with the $IPR^{n}$ are also displayed. One can see that the bulk eigenstate can be more localized than that of the zero-energy edge state (Figs. \ref{fig3}(d$_{1}$) and \ref{fig3}d($_{3}$)). To directly visualize the different localization phenomena of system, we plot the spatial distribution of every wave function (including the bulk and edge) in Figs. \ref{fig3}(a$_{3}$-d$_{3}$) and bulk eigenstates solely  for different eigenenergies in Figs. \ref{fig3}(a$_{2}$-d$_{2}$), where eigenstates have been normalized. It can be shown that all bulk eigenstates spread out in the open chain in Figs. \ref{fig3}(a$_{2}$-a$_{3}$). Furthermore, some extended eigenstates can be transformed into the localized ones in Figs. \ref{fig3} (b$_{2}$-b$_{3}$). With the change of $W_{1}$, all extended eigenstates become localized in Figs. \ref{fig3} (c$_{2}$-c$_{3}$). However, some already localized eigenstates will become extended again in Figs. \ref{fig3} (d$_{2}$-d$_{3}$) upon further increasing the quasidisorder. Eventually, the bulk will revert to the localized again.

Finally, we consider the effect of the non-Hermiticity on the localization behaviors of the bulk and edge, respectively. In Fig. \ref{fig4}(a), the bulk exhibits two coexistence regimes with changing $\gamma$. Namely, the reentrant localized bulk can be reconstructed by increasing the non-Hermiticity. Note that the values of $NPR^{B}$ in the second coexistence area are small. To exclude the finite size effect, the detailed values of $NPR^{B}$ at $\gamma=4.66946$ with $L=20000$, $30000$ and $40000$ are detailed in Fig. \ref{fig4}(b). Obviously, $NPR^{B}$ remains nonzero in the thermodynamic limit. Fig. \ref{fig4}(c) shows the localization phenomena of the wave functions with the lowest energy. Clearly, the wave functions are localized whether the system is nontrivial or not except for at the localized phase transition point, or equivalently, at the topological phase transition point. It can be shown that the variation range of $NPR^{0}$ around the phase transition point is very small. Therefore, we effectively plot the derivative of $NPR^{0}$ in Fig. \ref{fig4}(d) with $L=12000$ once more, a steep change can be found in the phase transition points as mentioned above.

\section{CONCLUSION and discussion}\label{V}

To summarize, we have analyzed localization behaviors of the bulk and topological edge state based on a quasiperiodic non-Hermitian system, respectively. Our results reveal that the bulk can hold the reentrant localization phenomenon. Meanwhile, the numerical results reveal that the topological edge state can vanish and re-emerge with increasing the quasidisorder strength, and the degree of localization of the edge state is not positively correlated with the strength of quasidisorder. The points at which the topological phase transition occurs must correspond to those for the localized transition of the edge state. In addition, a sharp discontinuity in the derivative of the normalized participation ratio occurs at the phase transition point.

Although we take the simple Hamiltonian \eqref{2.1} as an example in this paper, the analysis processes (the definition of the topological invariant, and the separate discussion of the bulk and edge) and results can be extended to other systems in the framework of sublattice symmetry naturally. In addition, many techniques have been used to produce the non-Hermitian and quasidisorder [\onlinecite{Longhi53, Seba54, Kai55, Helb56, Gha57, Liu58, Bro59}], and our system could be realized in some artificial systems. The methodology adopted here may help deepen the understanding of the localization behaviors of the system.

\section{ACKNOWLEDGMENTS}
This work was supported by National Natural Science Foundation of China (Grants No. $11874190$, No. $61835013$ and No. $12047501$), and National Key R\&D Program of China under grants No. 2016YFA0301500. Support was also provided by Supercomputing Center of Lanzhou University.

\bibliography{paper}

\end{document}